\RequirePackage{fix-cm}
\documentclass[twocolumn]{svjour3}
\usepackage{graphics}
\smartqed  
\usepackage{graphicx}
\usepackage[colorlinks]{hyperref}
\usepackage{dcolumn}
\usepackage{amssymb}
\usepackage{amsmath}
\usepackage[usenames]{color}
\usepackage{widetext}
\newcommand{\ar}{\arrowvert}
\newcommand{\ra}{\rangle}
\newcommand{\la}{\langle}
\newcommand{\da}{\dagger}
\newcommand{\ov}{\overline}
\newcommand{\cd}{\! \cdot \!}
\newcommand{\be}{\begin{equation}}
\newcommand{\ee}{\end{equation}}
\newcommand{\ba}{\begin{eqnarray}}
\newcommand{\ea}{\end{eqnarray}}

\begin{document}
\title{A first estimate of triply heavy baryon masses \\
from the pNRQCD perturbative static potential}
\author{Felipe J. Llanes-Estrada, Olga I. Pavlova, Richard Williams 
}                     
\institute{F. J. Llanes-Estrada, O.I. Pavlova and R. Williams \at Departamento de F\'{\i}sica Te\'orica I,  Universidad
Complutense, 28040 Madrid, Spain 
}

\date{Received: date / Revised version: date}
%
\authorrunning{Llanes-Estrada et al.}
\titlerunning{Triply heavy baryons in pNRQCD}
\maketitle

\begin{abstract}
Within pNRQCD we compute the masses of spin-averaged triply heavy
baryons using the now-available NNLO pNRQCD potentials and three-body variational
approach. We focus in particular on the role of the purely three-body
interaction in perturbation theory. This we find to be reasonably small
and of the order 25~$MeV$
\\
Our prediction for the $\Omega_{ccc}$ baryon mass is $4900(250)$ in
keeping with other approaches. We propose to search for this hitherto unobserved state at 
$B$ factories by examining the end point of the recoil spectrum against triple charm. 
\end{abstract} 
%

\section{Introduction}
Amongst the staples of hadron physics is Baryon spectroscopy. Here,
quark model computations of the light baryon
spectrum~\cite{Melde:2008yr,Loring:2001kx,Glozman:1994gt} find only mild
success beyond ground-states in various channels due to the plethora of
open thresholds and couplings between channels. A much cleaner system
can be provided by baryons composed of three heavy quarks (i.e. the
combinations $ccc$, $bbb$, $ccb$ and $bbc$). Since a considerable number
of states are supposed to be below any strong decay thresholds, one can
straightforwardly apply few-body reasoning, and quark
model techniques can handle bound states better.

Indeed, a review of the literature reveals many studies that have computed
or constrained heavy baryon masses, particularly the ground state
$\Omega_{ccc}$. We collect some of the values obtained in
section~\ref{PriorComps} below. Like its meson (quarkonium)
counterpart~\cite{Brambilla:2010cs,Brambilla:2004wf}, we expect this
triply heavy baryon to attract much interest.

For heavy quark systems, the development of potential Non-Relativistic Quantum Chromodynamics
(pNRQCD) as an effective theory of QCD has allowed a
more systematic treatment of the theoretical uncertainties involved in
spectroscopic predictions by expanding in powers of
$1/m$~\cite{Brambilla:2004jw,Brambilla:2000gk,Pineda:2000sz}. For those
in their ground state, pNRQCD can additionally be organized in standard
perturbation theory as a power expansion in
$\alpha_s$~\cite{Brambilla:2004jw,Brambilla:1999xf}.    While the
theory itself has limitations due to the finiteness of the quark masses, the
two-body static potential (quickly reviewed in
section~\ref{2bodypotential}) has shown to be a good starting
point for many meson investigations.

With the three-body potential in NNLO perturbation theory now at
hand~\cite{Brambilla:2009cd} (we will give it in section~\ref{3bodypot}), 
it is timely to perform an exploratory study of the ground-state
triply heavy quark spectrum. This we present in
section~\ref{3bodycomputation}.
The necessary QCD parameters $\alpha_s$, $m_c$, $m_b$ are fixed by
describing several common meson spectroscopy observables as explained in
section~\ref{observables}.

Finally, we comment on the feasibility of detecting the $\Omega_{ccc}$ in
section~\ref{prospects}. Some numerical methods are relegated to the
appendix.

\section{Static quarkonium potential in pNRQCD}\label{2bodypotential}
%
The static two-body potential for bound states of a quark and anti-quark is well
known to NNLO (and
beyond)~\cite{Vairo:2000ia,Schroder:1998vy,Peter:1997me} as
\begin{align}\label{2body}
V^{(0)} &= V_{LO}^{(0)} + V_{NLO}^{(0)} + V_{NNLO}^{(0)} + \cdots\,\, .
\end{align}
The leading order potential is just the color Coulomb potential
\begin{align}
V_{LO}^{(0)} &= -\frac{4}{3} \frac{\alpha_s(r^{-2})}{r}
\end{align}
whereas the NLO and NNLO are, respectively,
\begin{align} 
V_{NLO}^{(0)} &= V_{LO}^{(0)}\times \frac{\alpha_s(r^{-2})}{4\pi}
		     \times (a_1+2\gamma_E\beta_0) \,\, , \\
	\label{2bodypots}	
V_{NNLO}^{(0)}& =V_{LO}^{(0)}\times \frac{\alpha_s^2(r^{-2})}{(4\pi)^2}
\nonumber\\[-3mm] \\[-3mm] \nonumber
&\times\left( \gamma_E (4a_1\beta_0+2\beta_1) +
(\frac{\pi^2}{3}+4\gamma_E^2)\beta_0^2 + a_2\right)  \,\, .
\end{align}
By standard convention $\gamma_E \simeq 0.57721\dots$ is the
Euler-Mascheroni constant; for three colors and in terms of the number
of quark flavors $N_f$ below the renormalization scale, the beta
function that determines the running of the coupling is expanded as
\begin{align}
\beta_0 &\equiv 11- \frac{2N_f}{3}   \,\, , \\
\beta_1 &\equiv 102 - \frac{38N_f}{3}\,\, ,
\end{align}
and the coefficients in the potential that remain in a conformal theory are
\begin{align}
a_1&\equiv \frac{31}{3} - \frac{10N_f}{9} \,\, , \\
a_2&\equiv \frac{4343}{18} + 36 \pi^2-\frac{9\pi^4}{4} 
\nonumber \\[-3mm] \\[-3mm]  \nonumber
 +&66\zeta(3) -\left(\frac{1229}{27}+\frac{52}{3}\zeta(3)\right) N_f
 +\frac{100}{81} N_f^2 \,\, ,
\end{align}
where $\zeta(3)\simeq 1.202\dots$ is Riemann's zeta function.

The NLO potential can be understood not only from pNRQCD (first,
construct an effective theory around the heavy quark limit, then
use Coulomb gauge in intermediate steps to obtain the gauge-invariant
potential that is a matching coefficient in pNRQCD) but also from the
$D_{\sigma\sigma}(\ar \bf q \ar)$ time-like gluon 
propagator obtained in Coulomb gauge by Watson and
Reinhardt~\cite{Watson:2007mz}, reducing it to its simpler Heavy Quark
limit (there, several terms do not contribute to the Wilson
loop potential)\footnote{We thank Jacopo Ghiglieri for
this observation}.

In momentum space, the potential can be given to order NNLO~\cite{Schroder:1998vy}
at an arbitrary renormalization scale $\mu^2$ as
\begin{align}
	V(q^2) &= - \frac{4}{3}\frac{(4\pi)\alpha_{\ov{MS}}}{\vec{q}^2}\times \\ \nonumber
&\left(1+\frac{\alpha_{\ov{MS}}}{4\pi} C_1^{\ov{MS}}\left(\frac{\mu^2}{q^2} \right)
+\left( \frac{\alpha_{\ov{MS}}}{4\pi}\right)^2 C_2^{\ov{MS}}\left(\frac{\mu^2}{q^2}\right)
\right)
\end{align}
with
\begin{align*}
C_1^{\ov{MS}}(x) &= a_1 +\beta_0\log(x)     \,\, ,  \\ 
C_2^{\ov{MS}}(x) &= a_2 +\beta_0^2\log^2(x) +(\beta_1+2\beta_0
a_1)\log(x)\,\, .
\end{align*}
This obviously simplifies if the renormalization scale is chosen as $q^2$ itself
\begin{align} \label{qpotential}
V(q^2) =& -\frac{4}{3}\frac{(4\pi)\alpha_{\ov{MS}}(q^2)}{\vec{q}^2} \\ \nonumber
&\times\left( 1+a_1 \frac{\alpha_{\ov{MS}}(q^2)}{4\pi} 
+a_2 \left( \frac{\alpha_{\ov{MS}}(q^2)}{4\pi}\right)^2
\right) \,\, .
\end{align}

\subsection{$1/m$ Potential}\label{subsec:1overM}
%
In QED the $1/m$ corrections to the static potential were long ago shown
to vanish\footnote{For example, in~\cite{Ciafaloni:1968} it is shown
that the Bethe-Salpeter ladder approximation generates Feynman gauge
$1/m$ terms that vanish upon including the crossed-ladder box. No such
terms are present in Coulomb gauge.}.

In perturbative QCD the non-Abelian vertex correction produces a $1/m$
potential that cannot be gauged away. It can be nominally assigned to
the $1/m^2$ order through a field redefinition~\cite{Brambilla:2000gk}.
Since a recent study of the meson spectrum~\cite{Laschka:2011zr} finds
reasonably large effects for the $1S$ states, especially in charmonium,
we also consider it here.

At leading order, the $1/m$ potential vanishes. For NLO and NNLO we
employ the convention of~\cite{Kniehl:2001ju}. Alternatively, one can
use the NLO result (see Eq.~(\ref{1overmpot}) below)
without~\cite{Brambilla:2000gk} the factor $(7/9)$, in order to match a
lattice computation.

In coordinate space the potential reads
\begin{align}
V_{m^{-1}} &= -\frac{\alpha_s^2(\mu)}{m_r r^2} \times \left(\frac{7}{9}\right) 
\nonumber \\[-5mm]\label{1overmpot} \\ \nonumber
&-\frac{\alpha_s^3(\mu)}{3\pi m_r r^2}
\left\{
-b_2 +  \log(e^{2\gamma_E}\mu^2 r^2)
\left( \frac{7\beta_0}{6}+\frac{68}{3} \right) \right\},
\end{align}
where $m_r$ is the reduced (pole) mass of the $q\bar{q}$ system,
$b_2\simeq -20.836$ for $N_f=3$ (appropriate for charmonium) and
$b_2\simeq -18.943$ for $N_f=4$ (appropriate for bottomonium) and
$b_2\simeq -17.049$ for higher scales where $N_f=5$ are given
in~\cite{Kniehl:2001ju}. The last term with a logarithm vanishes if the
scale is chosen as the BLM scale defined by Eq.~(\ref{BLMscale}) in the
appendix. We have performed computations with both this running scale and a fixed
scale ($m_c^2$ or $m_b^2$).

If the potential is constructed in momentum space, the $1/m$ correction to the central static potential reads
\begin{align}
V_{m^{-1}} &=
-2\pi^2 \frac{\alpha_s^2(\mu^2)}{m_r q} \times \left(\frac{7}{9}\right)
\nonumber \\[-5mm]\\ \nonumber
&-\frac{2\pi\alpha_s^3(\mu^2)}{3m_r q}\left(
-b_2 + \log\left( \frac{\mu^2}{q^2}\right)\left[
\frac{7\beta_0}{6}+\frac{68}{3}
\right]
\right) \,\, .
\end{align}
Again, judicious choice of the scale $\mu=q$ disposes of the logarithm.
 
A counterintuitive result is that matrix elements of the $V_{m^{-1}}$
potential can actually be similar or, in extreme cases, 
even larger for bottom systems than for charm systems, 
since in a Coulombic system all energies scale with
the reduced mass $m_r$.

To see this, let us restrict ourselves to NLO and employ the convention
of~\cite{Brambilla:2000gk} in momentum space
\begin{align}
\label{1overmpot2}
V_{m^{-1}} &= -2\pi^2 \frac{\alpha_s^2(\mu^2)}{m_r q}\,\, ,
\end{align}
and compute $\la \psi \ar V_{m^{-1}} \ar \psi \ra$ with a hydrogen like
wave-function. Taking the Fourier transform of a $1S$ state
$2e^{-r/a_0}/\sqrt{4\pi a_0^3}$, with Bohr radius $a_0^{-1}=m_r\alpha_s$
yields
\begin{align}
\psi(q) &= \frac{4\sqrt{4\pi a_0^3}}{(1+q^2a_0^2)^2} \,\, ,
\end{align}
and therefore
\begin{align}
\la V_{m^{-1}}\ra &= \frac{-2a_0^3}{m_r\pi^3}\int \frac{d^3kd^3q}{\ar
\vec{k}-\vec{q}\ar}  \frac{\alpha_s^2(\ar \vec{k}-\vec{q}\ar)}{\left((1+q^2a_0^2)
(1+k^2a_0^2)\right)^2} \,\, .
\end{align}
Extracting the dimensions and coupling, substituting the Bohr radius,
and not minding about two constant positive numerical coefficients $c$,
$c'$, we find
\begin{align} \label{1overMsurprise}
\la V_{m^{-1}}\ra &= -m_r \alpha_s^4(\alpha_s m_r c) c' \,\, .
\end{align}
If the coupling constant did not run, the expectation value (though
suppressed in the perturbative counting) would be some factor of
$m_b/m_c\simeq 3$ larger for bottom than for charm systems. The NLO
running of $\alpha_s$ however tames this growth and we will find very
modest increases as a function of quark mass between charm and bottom.

To give an example let's take pole quark masses of $m_c=1.95\ GeV$, $m_b=5.14\ GeV$ and $\alpha_s(m_Z)=0.114$ (various fits of these quantities will be given later on in section~\ref{fits}, these serve as illustration).

 At the soft scale, with reduced quark mass $\frac{m}{2}$, solving iteratively as shown in appendix~\ref{app:iterative}, 
$\alpha_s \left( \frac{m_c}{2} \alpha_s \right)=0.625$
and $\alpha_s \left( \frac{m_b}{2} \alpha_s \right)=0.395$   
Then $\alpha_s^4(c)/\alpha_s^4(b)\simeq 6.3$. If the constant $c$ in Eq.~(\ref{1overMsurprise}) is somewhat smaller than one, this number could be smaller and around 3.

\subsection{Running of the strong coupling constant}

The renormalization group equation that determines the running of the
strong coupling constant to NNLO is
\begin{align} \label{RGEcoupling}
\frac{\partial \alpha_s}{\partial \log q^2}&= 
-\frac{\beta_0}{4\pi}\alpha_s^2 - \frac{\beta_1}{(4\pi)^2}\alpha_s^3
\,\, .
\end{align}
By keeping the first term on the right hand side, or both terms, this
equation can be solved to NLO or NNLO respectively.

\begin{figure}
\centering
\includegraphics*[width=0.95\columnwidth]{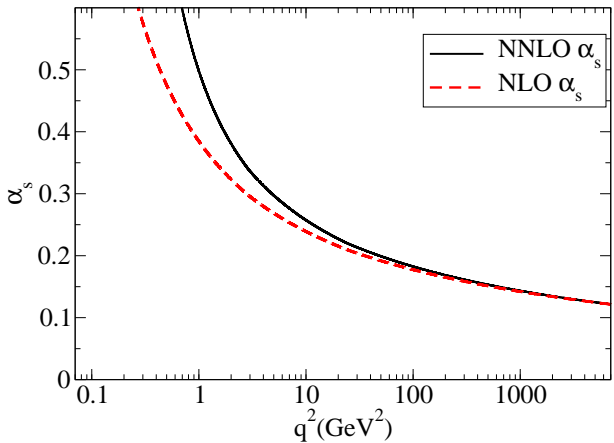}
\caption{\label{fig:alphasZ} Typical running coupling constant to order
NNLO agreeing with recent $\tau$ data and the world
average~\cite{Bethke:2009jm}.}
\end{figure}

In general we employ the Runge-Kutta algorithm to numerically solve Eq.~(\ref{RGEcoupling}).
To NLO the equation is also very simply analytically solvable and provides a handy check
for the computer programme. Following~\cite{Altarelli:2002wg} we
introduce a scale $\Lambda$ as is customary, so that
\begin{align}
\alpha_s^{NLO}(Q^2) &= \frac{1}{b\log \frac{Q^2}{\Lambda^2}} \,\, ,
\end{align}
with 
\begin{align*}
b= \frac{\beta_0}{4\pi} = \frac{33-2N_f}{12\pi} \,\, .
\end{align*}
Inverting the equation yields
\begin{align} \label{LambdaQCD}
\Lambda^2 &= Q^2 \exp{\left(-\frac{1}{b\alpha_s^{NLO}(Q^2)}\right)} \,\ .
\end{align}
In Table~\ref{tab:runningalpha} we give for convenience, and as
benchmarks, the values obtained by running back to low scales the renormalization group
equation from the Z-boson pole, where the coupling constant is very
accurately constrained by many analyses~\cite{Bethke:2009jm}. 
At each of the scales mentioned in the table the number of
active flavors in the beta function is decreased by one in a stepwise
fashion and continuous matching is performed\footnote{This introduces a
non-analyticity that can, at least at NLO, be avoided by the use of the
Brodsky-Lepage-Mackenzie method. See appendix~\ref{sec:BLM}. This non-analyticity is enhanced if one employs the discontinuous matching conditions~\cite{hep-ph/9305305} based on effective theory, that we also intend to incorporate in future work.}. A
typical run in agreement with world average and low-scale $\tau$ data is
plotted in figure~\ref{fig:alphasZ}. 
\begin{table}
\caption{Benchmarks for $\alpha_s(\mu^2)$ at various scales. Input is
the value of the coupling at the $Z$ mass~\cite{Bethke:2009jm}. Shown
are NLO and NNLO results with the number of active flavors decreased in
a step-wise fashion from five to four and then three at each of the
benchmark thresholds.\label{tab:runningalpha}.}
\begin{center}
\begin{tabular}{|cc|cc|} \hline
$\mu$   &  $N_f$ &  NLO  & NNLO \\ \hline
91.188  &  5     &  0.1184(7) & 0.1184(7) \\
5       &  4     &  0.204(2)  & 0.2136(6) \\
1.6     &  3     &  0.295(4)  & 0.336(2)  \\
0.8     &  -     &  0.417(8)  & 0.574(7)  \\
\hline
\end{tabular}
\end{center}
\end{table}
In addition we employ a second coupling that provides a best fit to key
charmonium and bottomonium data, that while still broadly consistent
with high energy data, is somewhat smaller. 

Moreover, we vary the number of flavors dynamically in the computer
programme upon crossing each threshold. Since the quark masses
themselves are varying during each fit we cannot quote these thresholds
here. However, they are in the vicinity of $1.6$/$1.7$~$GeV$
for the charm and $5$~$GeV$ for the bottom thresholds.

We know that the value of $\alpha_s$ at the Z-boson pole, evolved by
backward NNLO running, is consistent with the very precise determination
based on radiative decays of the $\Upsilon$
meson~\cite{Brambilla:2007cz}, $0.184\pm 0.015$ at the bottomonium
scale. Therefore, deviations from the Z-pole value in our fit results
give us an idea of the size of the errors associated with the
heavy-quark effective potentials. No further perturbation-theory approximation is
implicit, given that the Schroedinger equation is exactly diagonalized.

Finally, we remark that infinitely heavy quarkonium should be
less sensitive to the infrared details of the interaction, but that some
sizeable sensitivity remains because the quark mass is finite. Thus we employ the
heavy quark-potential for all $r$ in the programme. To avoid
encountering spurious Landau pole singularities, we freeze the coupling
constant at a low scale (400-600 $MeV$) and check the sensitivity to
this procedure below.

\begin{figure}
\centering
\includegraphics*[width=0.95\columnwidth]{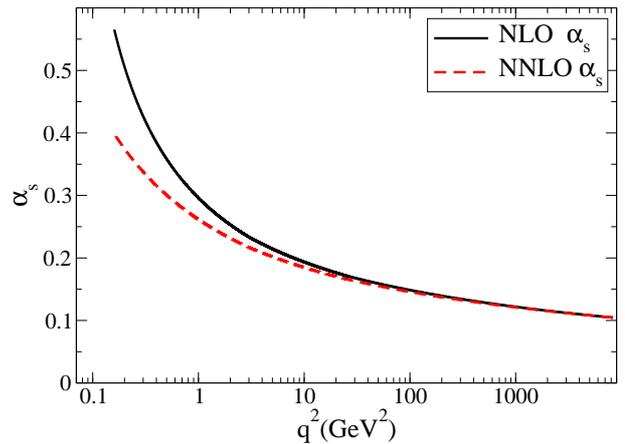}
\caption{\label{fig:alphas}Typical running coupling fit to two-body data. 
The actual best fit scale will depend
on the order of perturbation theory, possible renormalon subtraction and
treatment of the infrared, whether the static potential is or is not
corrected by the $1/m$ force, etc., and varies from computation to
computation as will be indicated below.}
\end{figure}

In conclusion, each of the runs reported will employ a slightly
different fit value of $\alpha_s$, but typical shapes for the coupling
constant can be seen in figures~\ref{fig:alphasZ} and \ref{fig:alphas}.

\section{Heavy baryon potential in perturbation theory}\label{3bodypot}
%
The static potential between three heavy quarks of equal mass in
positions $\vec{r}_1$, $\vec{r}_2$, $\vec{r}_3$ is expanded in powers of
the strong coupling constant as
\begin{align}
V^{(0)}(\vec{r}_1, \vec{r}_2, \vec{r}_3) &= V^{(0)}_{LO} + V^{(0)}_{NLO}
+ V^{(0)}_{NNLO}+ \cdots  \,\, .
\end{align}
In this work we will obtain the masses of a few heavy baryons from the
Leading Order and Next to Leading Order potentials by employing a
variational basis. Then we will study the effect of the intrinsic
three-body force (star-shaped) that appears first at NNLO, and finally
employ the rest of the known two-body NNLO terms to estimate further
corrections to the mass values.

The Leading Order potential is $\Delta$-shaped i.e. given by the sum of
the two-body Coulomb interactions
\begin{align} \label{LObaryonpotential}
V^{(0)}_{LO} &= \frac{-2\alpha_s}{3}\left(
\frac{1}{\ar \vec{r}_1-\vec{r}_2\ar}+
\frac{1}{\ar \vec{r}_2-\vec{r}_3\ar}+
\frac{1}{\ar \vec{r}_3-\vec{r}_1\ar} \right) \,\, .
\end{align}
For the remainder of this section we will shorten the notation by
summing over an index $i=1,2,3$ that runs over the three possible
pairings of the quarks, so that
\begin{align}
V^{(0)}_{LO} &= \frac{-2\alpha_s}{3} \sum_i \frac{1}{\vec{r}_i} \,\, .
\end{align}

The one-loop corrections to this potential yield the NLO part. The
coupling constant is renormalized and one needs to choose the
renormalization scale at which the constant is initially given.
Following the pNRQCD custom, we first select the renormalization scale
$\mu_i^2=1/{\ar \vec{r}_i \ar^2}$. Then the potential to
NLO~\cite{Brambilla:2009cd} reads
\begin{align}
V^{(0)}_{LO}+ V^{(0)}_{NLO} =&  
-\frac{2}{3} \sum_i \alpha_s(\ar \vec{r}_i \ar^{-2}) \frac{1}{\ar \vec{r}_i\ar } \times \\ \nonumber
& \left[ 1 + \frac{\alpha_s(\ar \vec{r}_i \ar^{-2})}{4\pi}
\left(2\beta_0\gamma_E + a_1
\right)\right]\,\, .
\end{align}

In momentum space, the NLO potential is easily reconstructed by
comparison with Eq.~(\ref{qpotential}).

We now turn to the potential at NNLO. We are first of all interested in the intrinsic three-body piece, the star-shaped part of the potential that appears at this order. While three-body forces have been considered in the context of heavy hybrid mesons~\cite{Guo:2007sm}, applications to triply heavy baryons are sparse.

\begin{figure}
	\centering
\includegraphics*[width=0.95\columnwidth]{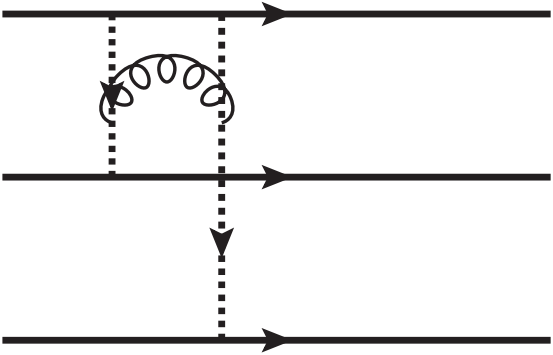}
\caption{One of twelve Feynman diagrams contributing to the intrinsic three-body force in a triply heavy baryon at NNLO.}
\end{figure}

This three-body force is conveniently organized in terms of an
auxiliary potential
\begin{align}
V_{NNLO-3}^{(0)} = 2\big( V_{\rm aux}({\bf r}_2,{\bf r}_3)
                  &+V_{\rm aux}({\bf r}_1,-{\bf r}_3) \nonumber \\
		     &+V_{\rm aux}(-{\bf r}_2,-{\bf r}_1) \big) \,\,,
\end{align}
that is computed via the Fourier transform 
\begin{align}
V_{\rm aux}({\bf r}_2,{\bf r}_3) = i\int\frac{d{\bf q}_2d{\bf q}_3}{(2\pi)^6}
e^{i{\bf q}_2\cdot {\bf r}_2} e^{i{\bf q}_3\cdot {\bf r}_3}
\hat{V}_{\rm aux}({\bf q}_2,{\bf q}_3)
\end{align}
of a potential
\begin{align} \label{3bodyforce}
\hat{V}_{\rm aux}(&{\bf q}_2,{\bf q}_3)= \frac{(-i/2)(4\pi)^3 \alpha_s^3}
{8\ar {\bf q}_2\ar^2 \ar{\bf q}_3\ar^2} \times \\ \nonumber
&\left[
\frac{\ar {\bf q}_2+{\bf q}_3\ar}{\ar {\bf q}_2 \ar\ar {\bf q}_3 \ar}
+\frac{{\bf q}_2\cd {\bf q}_3 + \ar {\bf q}_2 \ar\ar {\bf q}_3 \ar}
{\ar {\bf q}_2 \ar\ar {\bf q}_3 \ar \ar {\bf q}_2+{\bf q}_3\ar}
-\frac{1}{\ar {\bf q}_2 \ar}  -\frac{1}{\ar {\bf q}_3 \ar}  
\right]\ .
\end{align}
The two momenta $q_2$ and $q_3$ are flowing out of the two quark lines. 

The 2-body ($\Delta$-like) contribution to the potential at NNLO is
a simple generalization of Eq.~(\ref{2bodypots})
\begin{align}
V_{NNLO-2}^{(0)} =&-\frac{2}{3} \sum_i \frac{\alpha_s({\bf r}_i^{-2})}{\ar {\bf r}_i\ar} \frac{\alpha_s({\bf r}_i^{-2})^2}{(4\pi)^2}\times
\\ \nonumber
&\bigg( a_2 -36\pi^2+3\pi^4 +\left( \frac{\pi^2}{3} + 4\gamma_E^2\right)
\beta_0^2 \nonumber \\
&+\gamma_E(4a_1\beta_0+2\beta_1)
\bigg) \,\, .
\end{align}

That ground state heavy baryons feel more of a two-body like $\Delta$-shaped 
rather than the $\Upsilon$-shaped potential is supported by lattice
data, where the $\Delta$ ansatz seems dominant up to distances of
$R\simeq 0.7\ fm$~\cite{Alexandrou:2002sn} or even $1\ fm$~\cite{Bakry:2011ew}.

\section{Infrared subtracted potential}

The static potential in terms of the pole mass 
is afflicted by an infrared singularity in perturbation theory
~\cite{Brambilla:1999qa,Hoang:1998uv,Pinedasthesis} that can be
subtracted by a momentum-space cutoff~\cite{Beneke:1998rk}.
In passing from momentum space 
to coordinate space this amounts to a restricted Fourier transform
\begin{align} \label{subtPot}
V_{PS}(r,\mu_f) = \int_{\ar {\bf q}\ar> \mu_f} \frac{d^3q}{(2\pi)^3}
e^{i{\bf q}\cd {\bf r}} V(\ar {\bf q}\ar) \,\, .
\end{align}
Correspondingly, the quark mass one deduces from the fits to data is in this
particular ``Potential Subtracted'' scheme (shortened ``PS'' in what
follows). The infrared singularity (and accordingly, bad behavior of
perturbation theory) is now exposed as a counterterm necessary if one
wants to retrieve the pole mass
\begin{align}
m_{PS}(\mu_f) = m + \frac{1}{2}\int_{\ar {\bf q}\ar< \mu_f}  
\frac{d^3q}{(2\pi)^3} V(\ar {\bf q}\ar)\ .
\end{align}
In perturbation theory, lacking another scale except the renormalization
scale, the counterterm has to be proportional to $\mu_f$ up to
logarithms, and is displayed explicitly in Eq.~(\ref{PStopole}) below.
This is so that the static energy of the system does not depend on the subtraction, 
\begin{align}
V_{PS} + 2m_{PS} = V + 2 m \,\, .
\end{align}
One can avoid the singularity altogether by directly relating the $PS$
mass with the $\ov{MS}$ mass by means of Eqs.~(\ref{poletoMS}) and
(\ref{PStopole}).

The relation between pole quark masses $m$ and $\ov{MS}$ masses $\ov{m}$ 
to NNLO is given in terms of the number of light quarks $N_f$ as
\begin{align}\label{poletoMS} 
m &= \ov{m}(\ov{m}) \times \\ \nonumber
&\left[ 1 + \frac{4}{3}\frac{\alpha_s(\ov{m})}{\pi}
+\left(13.44 -1.041 N_f \right) 
\left( \frac{\alpha_s(\ov{m})}{\pi} \right)^2
\right]\, , 
\end{align}
where $\alpha_s$ is understood as the $\ov{MS}$ coupling constant (in
terms of which we have anyway expressed the heavy quark static
potential). The inverse relation is~\cite{Chetyrkin:1999ys}
\begin{align}
\ov{m}(m&) = m \times \\ \nonumber
&\!\!\!\!\left[ 1 - \frac{4}{3}\frac{\alpha_s(\ov{m})}{\pi}
+\left(-14.33 + 1.041 N_f \right) 
\left( \frac{\alpha_s(\ov{m})}{\pi} \right)^2
\right] . 
\end{align}

While using the Potential Subtraction scheme we need to convert the PS mass
$m_{PS}$ to the pole mass too. This is achieved by means
of~\cite{Beneke:1998rk}
\begin{align} \label{PStopole}
m=&m_{PS}(\mu_f)+ \mu_f C_F\frac{\alpha_s(\mu)}{\pi}\times \\ \nonumber
&\left( 1+\frac{\alpha_s(\mu)}{4\pi} 
     \left[ a_1 -\beta_0 \left( \log \left( \frac{\mu_f^2}{\mu^2}\right)-2
                        \right)
    \right]
\right)\,\, .
\end{align}

\section{Meson observables}\label{observables}

If we handle the heavy charmonium and heavy bottomonium systems we need
to fix three parameters, the strong coupling constant  at one scale
$\alpha_s(\mu^2)$ (the evolution equation of the renormalization group
then provides it at any other scale) and the quark masses at their pole
(or in any other scheme). Alternatively, we employ a fixed running
coupling constant from figure~\ref{fig:alphasZ} and fit only the quark
masses to the ground state quarkonia (thus limiting ourselves to two
parameters).

We will employ five observables to fix these parameters, two from
charmonium ($1S$ mass and radiative transition $J/\psi\to
\eta_c\gamma$), two from bottomonium ($1S$ and $1P$ masses), and the
$B_c$ mass. This will over constrain the parameters for additional
security and allow us to check the running of the coupling between the
two scales, at the charm and at the bottom.

Since the purely static potential does not accommodate a hyperfine
${\bf S}_1 \cd {\bf S}_2$, nor an ${\bf L}\cd{\bf S}$ splitting, we will
employ spin-averaged masses. For bottomonium, we employ the recently
measured $\eta_b$ mass, 9391(3) $MeV$, and the $\Upsilon$ mass
$9460.3(3)\ MeV$, whose spin average is
$(M_{\eta_b}+3M_{\Upsilon})/4=9443(1)\ MeV$. We also have at our
disposal the $P$-wave mesons, that are also sometimes believed to lie in
the regime where perturbative pNRQCD is applicable,
$\chi_{b0}$ with mass $9859.4(4)\ MeV$,  
$\chi_{b1}$ with mass $9892.8(3)\ MeV$, and
$\chi_{b2}$ with mass $9912.2(3)\ MeV$,
that yield a spin average $(M_{\chi_{b0}}+3M_{\chi_{b1}} +5M_{\chi_{b2}})/9 =9899.9(3)\ MeV$.

Looking towards the $2S$ bottomonium levels (that, although their
wave-function starts being immersed in the non-perturbative part of the
potential, can still to some approximation be considered perturbative
quarkonium states), we notice that the $\eta_b(2S)$ mass has not been
measured yet, so that we take the $\Upsilon(2S)$ as sole data, without
the possibility of spin-averaging. The error introduced is expected to
be of order $10\ MeV$ or less since the hyperfine splitting should be
smaller than for the ground state (where the difference between the
$\Upsilon$ and the spin average is $17\ MeV$). One shouldn't expect this
state to be well described in perturbation theory, but we show results
nonetheless for comparison and completeness.

Turning to charmonium, the renowned $J/\psi$ state has a currently
accepted mass of $3096.92(1)\ MeV$, and the $\eta_c(1S)$ a mass of
$2980(1)\ MeV$. Their spin averaged mass is therefore
$(M_{\eta_c}+3M_{J/\psi})/4=3067.7(3)\ MeV$. The NRQCD description of
the $2S$ states in the charmonium system is generally not accepted, so
to obtain one more observable at the charmonium scale we turn to the
$S$-wave radiative transition width $J/\psi\to \eta_c \gamma$, decay
branch $\Gamma_{133}$ of the $J/\psi$ in the notation of
\cite{Nakamura:2010zzi}. This branching fraction  currently yields a
width $\Gamma_{J/\psi\to \eta_c \gamma} = 1.6(4)\ keV$.

This radiative transition width between the $1S$ charmonium states has
also been calculated~\cite{Brambilla:2005zw} at NNLO, and all we need to
do is evaluate it numerically.

In addition we have at our disposal the mass of the $B_c$ meson,
$M_{B_c}=6277(6)\ MeV$. Here the vector $B_c^*$ is not yet known, so
that the spin average has to be obtained by interpolation from other
flavor combinations. Currently we can lean on the $B$ system for which
the hyperfine splitting is $\Delta M_{hf}=45.8(3)\ MeV$,  the $B_s$
system where $\Delta M_{hf}=49(2)\ MeV$, and the bottomonium
$\Upsilon-\eta_b$ system where $\Delta M_{hf}=69(3)\ MeV$. Our
interpolated estimate for the $B_c$ flavor combination is therefore
$\Delta M_{hf}=53(3)\ MeV$. Adding this correction to the $B_c$ mass to
yield a spin average $M_{B_c} + \frac{3}{4}\Delta M_{hf}$ gives
$6317(8)\ MeV$.

This parameter set is collected in table~\ref{tab:data}.
\begin{table}
\caption{Experimental data employed to fix the potential parameters
(adapted from \cite{Nakamura:2010zzi}). The error in parenthesis refers
to the last significant digit.\label{tab:data}}
\begin{center}
\begin{tabular}{|cc|} \hline
$(M_{\eta_b}+3M_{\Upsilon})/4$ & $9443(1)\ MeV$ \\
$(M_{\chi_{b0}}+3M_{\chi_{b1}} +5M_{\chi_{b2}})/9$ & $9899.9(3)\ MeV$\\
$M_{\Upsilon(2S)}$ & $10 023(0.3)\ MeV$ \\
$(M_{\eta_c}+3M_{J/\psi})/4$       & $3067.7(3)\ MeV$ \\
$\Gamma_{J/\psi\to \eta_c \gamma}$ & $1.6(4)\ keV$ \\
$M_{B_c}+(3/4)\Delta M_{hf}$ & $6317(8)\ MeV$ \\
\hline
\end{tabular}
\end{center}
\end{table}

\subsection{Evaluation of $\Gamma_{J/\psi\to \eta_c \gamma}$}
\label{subsec:rad}
Here we comment on the numerical evaluation of the radiative transition between the two $1S$ charmonium states, $J/\psi$ and $\eta/c$.
At NNLO we read off~\cite{Brambilla:2005zw}
\begin{align}
\Gamma_{J/\psi\to \eta_c \gamma} &= \frac{16\alpha_{\rm em}}{3} e_c^2
\frac{k_\gamma^3}{M^2_{J/\psi}} \times \\ \nonumber
&\left(
1+\frac{4}{3} \frac{\alpha_s((M_{J/\psi}/2)^2)}{\pi}-
\frac{2}{3}\left( \frac{4}{3}\alpha_s(r^{-2}) 
\right)^2 \right)\,\,,
\end{align}
where $e_c=2/3$ is the charm quark's charge in units of the electron
charge, and the pole quark mass has been eliminated in terms of the
physical $J/\psi$ mass. Since the experimental error in this observable
is relatively large, it is not sensible to delve into subtleties about
this elimination.

The momentum of the photon is given simply as $k_\gamma
=(M_{J/\psi}^2-M_{\eta_c}^2)/(2M_{J/\psi})$ by energy-momentum
conservation.

The strong coupling constant is evaluated at two different scales in
this formula, half the resonance mass (corresponding to the scale of the
quark mass, or hard scale) and the momentum scale of the quark inside
the resonance $p_{J/\psi}=r^{-1}$ (or soft scale). The first one is of
direct evaluation, but since the characteristic momentum in charmonium
is $\la p \ra \simeq m_r \alpha_s = \frac{m_c}{2} \alpha_s$, the second occurrence of the
coupling constant has a scale that depends itself on the coupling
constant, $\alpha_s\left(\frac{M_{J/\psi}}{4} \alpha_s \right)$.

To obtain $\alpha_s$ one then needs to employ
Eq.~(\ref{runningcoupling2}) recursively. The method is detailed in
appendix~\ref{app:iterative}.

\section{Meson numerical results and parameter fixing}\label{fits}
\subsection{Exploratory fits}
In this subsection we gain a feeling for the parameter values and
explore several alternatives. Table~\ref{tab:params} presents the result
of our fits numbered consecutively, not including the $B_c$ mass nor
including the $1/m$ potential.

\begin{table*}
\begin{center}
\caption{\label{tab:params}
Numerical results and fit parameters. Units are $MeV$ except
for the (1S) radiative decay width $\Gamma_{J/\psi\to\eta_c\gamma}$
which is quoted in $keV$. The reference experimental data for the
various fits can be found on table~\ref{tab:data}.
The quark masses quoted are the pole masses and the $\ov{MS}$ masses at $3\ GeV$ to facilitate comparison. 
\emph{Notes:} $^\da$ at the limit of the allowed fitting range; $^*$fixed from the world average value at the $Z$ boson pole; $^\&$ running coupling separately fit to the charm and bottom data. }
\begin{tabular}{|cc|ccccc|ccccc|} \hline
Fit & Order & $m_{c\bar{c}}$ & $\Gamma_{J/\psi\to\eta_c\gamma}$ & $m_{b\bar{b}}(1S)$ & $m_{b\bar{b}}(2S)$ & $m_{b\bar{b}}(1P)$ & $\alpha_s(m_Z^2)$ & $m_c$ & $m_b$ & $\ov{m}_c$ & $\ov{m}_b$ \\ \hline
1& LO   &3068 & 2.7& 9477& 9836& 9913& 0.128            & 1720       &4980 & 820 & 4600 \\
2& NLO  &3068 & 3.0& 9567& 9808& 9908& 0.111            & 1740       &5080 & 1160 & 4780\\
3& NNLO &3068 & 2.6& 9632& 9820& 9901& 0.095            & 1700       &5050 & 1330 & 4900\\ \hline
4& NNLO &3068 & 2.6& 9536& 9806& 9911& 0.099            & 1720       &5040 & 1340 & 4860\\
5& NNLO &3068 & 1.9& 9443& 9813& 9986& 0.1184$^*$       & 2000$^\da$ &5220 & 1350& 4970\\ \hline
6& NNLO &3070 & 2.4& 9630& 9830& 9910& 0.107/0.097$^\&$ & 2000$^\da$ &5060 & &   \\
7& NNLO &3067 & 2.4& 9443& 9548& 9558& 0.108            & 2000$^\da$ &5170 & 1530& 4970\\ 
8& NNLO &3067 & 2.6& 9690& 9824& 9906& 0.095            & 1700       &5050 & 1368& 4900\\ 
9& NNLO &3068 & 0.74&9443& 9685& 9525& 0.12             & 2188       &5381 & 1540 & 5140  \\ \hline
10&NNLO &3068 & 1.6 &9443& 9688& 9796& 0.111            & 1853       &5100 & 1340 & 4878\\
\hline
\end{tabular}
\end{center}
\end{table*}

Fits $1$, $2$ and $3$ show the consistency of perturbation theory, which is
quite reasonable. In these fits the quark masses come out consistently
in the $1.7\ GeV$ (charm) and $5\ GeV$ (bottom) ranges. Here
$\alpha_s(m_Z^2)$, left free, varies more significantly upon increasing
the order of perturbation theory, and is largely ensuring that the $1P$
bottomonium level is in good agreement with experiment.  In spite of
this, the radiative $\psi\to \eta_c$ transition width is outside its
experimental $2\sigma$ error band, and the splittings in the bottomonium
spectrum are significantly smaller than experiment. The smallness of
this splitting pulls the spin-averaged $b\bar{b}(1S)$ mass to higher
than physical values. \\
The first of these defects is common to $c\bar{c}$ approaches, and
within the present scheme it requires certain fine tuning. The second
problem is related to the fact that bottomonium excitations start being
sensitive to the linear part of the static potential, so that a purely
perturbative quarkonium description is not very precise.

Fits $4$ and $5$ show the insensitivity to changing the freezing scale
for the running coupling constant, that is in these fixed to $0.8\ GeV$,
whereas it is $0.5$ in all others. Comparing fits $3$ and $4$ for
example, we see that the value of this constant is irrelevant for all
purposes, except a marginal improvement in the bottomonium splittings.\\
Fit $5$ is different in that the coupling constant is not allowed to
vary, but fixed to the world average at the Z-pole. This larger value of
the coupling brings the transition width to better agreement with
experimental data. This is due to the NNLO contribution, negative, being
much enhanced. The charm quark is however pushed to the
limit of its variation band between $1$ and $2~GeV$ in the programme, and increases disagreement with other determinations.

In fits $6$ and following we return to a freezing scale of 0.5 $GeV$ but
change the way to compute the $\chi^2$ to be minimized. Instead of
employing the experimental error bands $\sigma_{i\ \ exp}^2$ for the
quarkonium masses in
\begin{align}
\chi^2=\sum \frac{(E^{th}-E^{exp})_i^2}{\sigma_{i\ \ exp}^2} \,\, ,
\end{align}
we adopt a common error band of $30\ MeV$ for all of them. This is in
recognition that theory errors for these observables are orders of
magnitude larger than experimental errors, and we want to check that the
experimental errors are not weighing the various states unduly in the
fit.

In fit $6$ we separately fit the two charmonium observables and the
three bottomonium observables, to ascertain the tension between them.
This is visible from the two different values obtained for the coupling
constant evolved to the Z pole, at the level of $10\%$. \\
In fit $7$ we leave the bottomonium $2S$ and $1P$ excitations out of the
$\chi^2$ formula to guarantee that the bottom quark mass is fixed to the
bottomonium spin-averaged ground state. We see that the width
$\Gamma_{J/\psi\to\eta_c}$ pulls the charm quark mass again to the limit
of our allowed variation band.

In fit $8$ instead we decouple $\Gamma_{J/\psi\to\eta_c}$ from the
minimization. This immediately relaxes back the charm quark mass.

In fit $9$ we use as input the spin-averaged ground state masses of
bottomonium and charmonium, together with the pseudodata
$\alpha_s(m_Z^2)=0.12$, that can be understood as a fit to the ratio of
radiative to total widths of the $\Upsilon$~\cite{Brambilla:2007cz}. At
the $\tau$ pole the running coupling is also in agreement with $\tau$
data, that suggests $\alpha_s(\tau)=0.330(25)$~\cite{Pich:2011bb}. The
radiative width is now below the experimental value, showing that with
fine tuning of the coupling constant it can be brought to the physical
value. However the $1P$ bottomonium state is now very far off, due to
the increased coupling, and the quark masses are far from other
determinations.

We proceed to fit $10$, in which we force the reproduction of the
precise experimental number for the $1S$ radiative width of 1.6 $keV$.
This happens for $\alpha_s(m_Z^2)=0.111$ at NNLO. As expected, this is
possible at the expense of losing agreement with the $1P$ mass.

Since the experimental error in the radiative width is so much larger
than the error in the $1P$ mass measurement, a best fit will try to
compromise by lowering the coupling constant in spite of this
deteriorating the computation of the width.

Overall it appears that to obtain a perfect value for
$\Gamma_{J/\psi\to\eta_c}$ requires a little fine tuning, and that in no
case is it possible to obtain an excellent fit to all five quantities
simultaneously.

\subsection{Extended fits}

In this subsection we include the $B_c$ mass and explore in addition the
effect of the $1/m$ potential. In all cases the coupling constant is
evolved to the $Z$ pole at NNLO (this should be for broad comparison and
not taken as a detailed prediction since higher orders of perturbation
theory should then be used).

Comparing tables~\ref{tab:fits0.6} and \ref{tab:fits0.61overM} we see
that the effect of the $1/m$ potential is modest, the fit preferring a
slightly lower coupling constant, and the $B_c$ mass being better
adjusted.

\begin{table}[h]\caption{Further fits in pole scheme.  Masses are in
	$MeV$, the radiative decay width $J/\psi\to \eta_c\gamma$ in
	$keV$. The infrared freezing of the running coupling constant
	occurs at 0.6 $GeV$. The recoil $1/m$ potential is not included,
	only the static potential.\label{tab:fits0.6}}
\begin{center}
\begin{tabular}{|c|ccc|c|}\hline
               & LO & NLO & NNLO & Expt. \\
Fit number     & 1 & 2 & 3 & \\ \hline
$m_{c\bar{c}}$ & 3068& 3068& 3068& 3068\\
$\Gamma_{J/\psi\to\eta_c\gamma}$ & 2.7& 3.1& 2.1& 1.6(4)\\
$m_{b\bar{b}}(1S)$ & 9458& 9447& 9480& 9443 \\ 
$m_{b\bar{b}}(2S)$ & 9820& 9777& 9769& 10023\\
$m_{b\bar{b}}(1P)$ & 9899& 9900& 9897& 9900\\
$m_{bc}(1S)$       & 5922& 6158& 6158& 6317\\
\hline
$m_c$ & 1720& 1770& 1850& \\
$m_b$ & 4970& 5040& 5120& \\
$\alpha_s$ & 0.128 & 0.116 & 0.111 &  \\ \hline
\end{tabular}
\end{center}
\end{table}

\begin{table}[h]\caption{As in table~\ref{tab:fits0.6} but with the
	recoil $1/m$ potential included\label{tab:fits0.61overM}.}
\begin{center}
\begin{tabular}{|c|ccc|c|}\hline
               & LO & NLO & NNLO & Expt. \\
Fit number     & 4 & 5 & 6 & \\ \hline
$m_{c\bar{c}}$ & 3068& 3068& 3068& 3068\\
$\Gamma_{J/\psi\to\eta_c\gamma}$ & 2.7& 3.0& 2.5& 1.6(4)\\
$m_{b\bar{b}}(1S)$ & 9443& 9445& 9488& 9443 \\ 
$m_{b\bar{b}}(2S)$ & 9775& 9788& 9772& 10023\\
$m_{b\bar{b}}(1P)$ & 9900& 9900& 9896& 9900\\
$m_{bc}(1S)$       & 6330& 6210& 6364& 6317\\
\hline
$m_c$ & 1740& 1740& 1820& \\
$m_b$ & 5040& 5000& 5110& \\
$\alpha_s$ & 0.102 & 0.106 & 0.105 &  \\ \hline
\end{tabular}
\end{center}
\end{table}

\begin{table}[h]\caption{As in table~\ref{tab:fits0.6} but in the PS
	scheme with the potential totally cutoff in the infrared at 0.6
	$GeV$\label{tab:fits0.6PS}.}
\begin{center}
\begin{tabular}{|c|ccc|c|}\hline
               & LO & NLO & NNLO & Expt. \\
Fit number     & 7 & 8 & 9 & \\ \hline
$m_{c\bar{c}}$ & 3068& 3068& 3068& 3068\\
$\Gamma_{J/\psi\to\eta_c\gamma}$ & 2.7& 3.0& 2.5& 1.6(4)\\
$m_{b\bar{b}}(1S)$ & 9462& 9443& 9443& 9443 \\ 
$m_{b\bar{b}}(2S)$ & 9818& 9792& 9789& 10023\\
$m_{b\bar{b}}(1P)$ & 9898& 9900& 9900& 9900\\
$m_{bc}(1S)$       & 5866& 6264& 6270& 6317\\
\hline
$m_c$ & 1740& 1930& 1710& \\
$m_b$ & 4970& 5230& 5020& \\
$\alpha_s$ & 0.133 & 0.112 & 0.106 &  \\ \hline
\end{tabular}
\end{center}
\end{table}
Tables \ref{tab:fits0.6PS} and \ref{tab:fits0.6PS1overM} then show the same calculation but in the PS scheme.
An interesting feature in these computations is seen in the last three
rows of table~\ref{tab:fits0.6PS1overM}. When the $1/m$ correction is
included, the PS scheme seems to be rather stable in going from LO to
NLO to NNLO, as the quark masses barely change.

\begin{table}[h]\caption{As in table~\ref{tab:fits0.6} but in the PS
	scheme with the potential totally cutoff in the infrared at 0.6
	$GeV$, and with the $1/m$ recoil potential included
	\label{tab:fits0.6PS1overM}.}
\begin{center}
\begin{tabular}{|c|ccc|c|}\hline
               & LO & NLO & NNLO & Expt. \\
Fit number     & 7 & 8 & 9 & \\ \hline
$m_{c\bar{c}}$ & 3068& 3068& 3068& 3068\\
$\Gamma_{J/\psi\to\eta_c\gamma}$ & 2.6& 3.0& 2.5& 1.6(4)\\
$m_{b\bar{b}}(1S)$ & 9443& 9444& 9443& 9443 \\ 
$m_{b\bar{b}}(2S)$ & 9791& 9792& 9789& 10023\\
$m_{b\bar{b}}(1P)$ & 9900& 9900& 9900& 9900\\
$m_{bc}(1S)$       & 6270& 6264& 6270& 6317\\
\hline
$m_c$ & 1690& 1700& 1710& \\
$m_b$ & 5000& 5020& 5010& \\
$\alpha_s$ & 0.106 & 0.112 & 0.106 &  \\ \hline
\end{tabular}
\end{center}
\end{table}

Comparing tables~\ref{tab:fits0.4} and \ref{tab:fits0.4_2} we see that,
with a lower infrared cutoff, the PS scheme does somewhat better in
terms of convergence and agreement with data.

\begin{table}[h]\caption{Further fits.  Masses are in $MeV$, the
	radiative decay width $J/\psi\to \eta_c\gamma$ in $keV$. The
	infrared freezing of the running coupling constant occurs at 0.4
	$GeV$.\label{tab:fits0.4}}
\begin{center}
\begin{tabular}{|c|ccc|c|}\hline
               & LO & NLO & NNLO & Expt. \\
Fit number     & 1a & 2a & 3a & \\ \hline
$m_{c\bar{c}}$ & 3068& 3068& 3068& 3068\\
$\Gamma_{J/\psi\to\eta_c\gamma}$ & 2.7& 3.1& 2.5& 1.6(4)\\
$m_{b\bar{b}}(1S)$ & 9542& 9481& 9577& 9443 \\ 
$m_{b\bar{b}}(2S)$ & 9829& 9800& 9804& 10023\\
$m_{b\bar{b}}(1P)$ & 9891& 9897& 9888& 9900\\
$m_{bc}(1S)$       & 6112& 6118& 6277& 6317\\
\hline
$m_c$ & 1667& 1828& 1761& \\
$m_b$ & 4963& 5117& 5084& \\
$\alpha_s$ & 0.120 & 0.121 & 0.104 &  \\ \hline
\end{tabular}
\end{center}
\end{table}

\begin{table}[h]\caption{Same as table \ref{tab:fits0.4} but in the PS
	scheme, with the potential completely cutoff in the infrared at
	0.4 $GeV$.\label{tab:fits0.4_2}}
\begin{center}
\begin{tabular}{|c|ccc|c|}\hline
               & LO & NLO & NNLO & Expt. \\
Fit number     & 7a & 8a & 9a & \\ \hline
$m_{c\bar{c}}$ & 3068& 3068& 3068& 3068\\
$\Gamma_{J/\psi\to\eta_c\gamma}$ & 2.7& 3.0& 2.5& 1.6(4)\\
$m_{b\bar{b}}(1S)$ & 9614& 9443& 9443& 9443 \\ 
$m_{b\bar{b}}(2S)$ & 9836& 9791& 9788& 10023\\
$m_{b\bar{b}}(1P)$ & 9884& 9900& 9900& 9900\\
$m_{bc}(1S)$       & 6218& 6265& 6269& 6317\\
\hline
$m_c$ & 1631& 1718& 1748& \\
$m_b$ & 4964& 5029& 5055& \\
$\alpha_s$ & 0.115 & 0.112 & 0.105 &  \\ \hline
\end{tabular}
\end{center}
\end{table}

Fixing the running coupling constant to either the world average or
recent $\tau$ data, as in tables \ref{tab:fitsfixedalpha} and
\ref{tab:fitsfixedalpha2} leads to slightly improved agreement with
experiment in the NNLO computation of the radiative transition width of
the $J/\psi$, but in exchange the $1P$ bottomonium mass is completely
off. 

\begin{table}[h]\caption{Further fits fixing the coupling constant at
	the Z-pole at $\alpha_s(m_Z)=0.1204$ as determined by recent
	$\tau$ data. Masses are in $MeV$, the radiative decay width
	$J/\psi\to \eta_c\gamma$ in $keV$. The infrared freezing of the
	running coupling constant occurs at 0.4 $GeV$.
\label{tab:fitsfixedalpha}}
\begin{center}
\begin{tabular}{|c|ccc|c|}\hline
               & LO & NLO & NNLO & Expt. \\
Fit number     & 1b & 2b & 3b & \\ \hline
$m_{c\bar{c}}$ & 3068& 3068& 3068& 3068\\
$\Gamma_{J/\psi\to\eta_c\gamma}$ & 2.7& 3.0& 0.65& 1.6(4)\\
$m_{b\bar{b}}(1S)$ & 9444& 9443& 9444& 9443 \\ 
$m_{b\bar{b}}(2S)$ & 9669& 9721& 9586& 10023\\
$m_{b\bar{b}}(1P)$ & 9718& 9812& 9654& 9900\\
$m_{bc}(1S)$       & 6145& 6195& 6149& 6317\\
\hline
$m_c$ & 1629& 1728& 1889& \\
$m_b$ & 4872& 5000& 5116& \\
$\alpha_s$ & & &  & fixed at 0.1204 \\ \hline
\end{tabular}
\end{center}
\end{table}

The difference between table~\ref{tab:fitsfixedalpha} and entry number 9
of table~\ref{tab:params} is that here the global fit strategy was used
while there only the $1S$ masses were used to constrain the quark
masses, and that the freezing of the coupling occurs at a slightly
different momentum (0.4 versus 0.46 $GeV$).

\begin{table}[h]\caption{Same as table~\ref{tab:fitsfixedalpha} but in
	the PS scheme, with the coupling constant cutoff at 0.4
	$GeV$.\label{tab:fitsfixedalpha2}}
\begin{center}
\begin{tabular}{|c|ccc|c|}\hline
               & LO & NLO & NNLO & Expt. \\
Fit number     & 7b & 8b & 9b & \\ \hline
$m_{c\bar{c}}$ & 3068& 3068& 3068& 3068\\
$\Gamma_{J/\psi\to\eta_c\gamma}$ & 2.7& 3.0& 0.65& 1.6(4)\\
$m_{b\bar{b}}(1S)$ & 9444& 9443& 9443& 9443 \\ 
$m_{b\bar{b}}(2S)$ & 9646& 9791& 9885& 10023\\
$m_{b\bar{b}}(1P)$ & 9690& 9900& 10034& 9900\\
$m_{bc}(1S)$       & 6155& 6264& 6248& 6317\\
\hline
$m_c$ & 1619& 1718& 1918& \\
$m_b$ & 4855& 5029& 5243& \\
$\alpha_s$ & & &  & fixed at 0.1204 \\ \hline
\end{tabular}
\end{center}
\end{table}

The charm quark mass has recently~\cite{Blossier:2010cr} been reobtained
from a lattice computation of the ground state charmed and charmonium
mesons ($D$, $J/\psi$, $\eta_c$), with $N_f=2$. In the $\ov{MS}$ scheme
they obtain
\begin{align}
m_c^{\ov{MS}}(2\ GeV) =1.14(4)\ GeV
\end{align}
that translates into a mass at the charm scale of
$$
\ov{m}_c(\ov{m}_c)= 1.28 (4) \ GeV \ .
$$
In comparing the quark mass between various schemes and
scales~\cite{Chetyrkin:1999pq}, the collaboration quotes an error less
than one standard deviation as a result of using $N_f=2$ instead of $N_f=4$.

Translating our pole-scheme masses in the various tables into $\ov{MS}$
masses $\ov{m}_c(\ov{m}_c)$ consistently yields results of order 1.3
$GeV$, in agreement with the lattice determination. The PS scheme on the
other hand gives a somewhat larger mass of about 1.4 $GeV$ in the
$\ov{MS}$ scheme.

To conclude this section let us quote an NLO computation with the BLM
scheme, with best fit $\alpha_s=0.113$, $m_c=1750\ MeV$ and
$m_b=5030\ MeV$. The $c\bar{c}$ radiative width is $3.0\ keV$. The $b\bar{b}$
(1S,2S,1P) masses are 9393, 9775 and 9904 $MeV$ respectively. The $B_c$
mass comes out to be 6180 $MeV$.  Thus, there is no particular advantage
in using this scheme over the PS or the pole ones.

\section{Prior computations of the baryon masses}\label{PriorComps}

\begin{figure*}
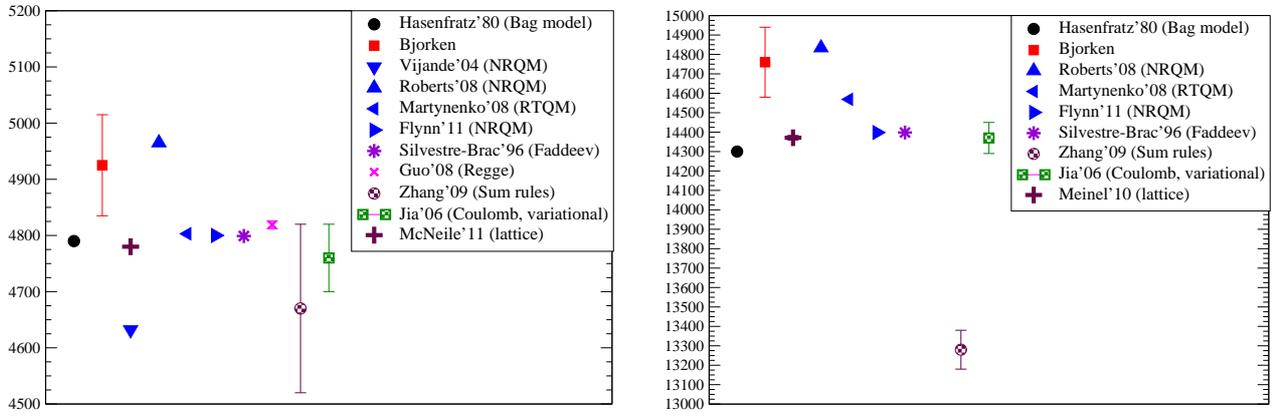

	\centering
\includegraphics*[width=8.cm]{FIGS.DIR/PriorComputationsccc.eps}\ \ \ \ \ \
\includegraphics*[width=8.cm]{FIGS.DIR/PriorComputationsbbb.eps}
\caption{Scatter of several existing computations for the $\Omega_{ccc}$
and $\Omega_{bbb}$ masses respectively. See table~\ref{tab:priorcomps}
for references.\label{fig:priorcomps}}
\end{figure*}

In this section we compile existing computations of the various ground state 
triply heavy baryons. Since we work in the static limit, we will not
resolve the hyperfine spin splitting between spin $1/2$ and spin $3/2$.
Thus, in table~\ref{tab:priorcomps} we quote the spin average
$(M_{1/2}+2M_{3/2})/3$ for the $bbc$ and $ccb$ wave-functions, that can
appear in both spin combinations in the
ground state.  We further plot some of these computations in
figure~\ref{fig:priorcomps}.

\begin{table*}
\caption{\label{tab:priorcomps}
Computations of triply heavy baryon masses. Many of the entries were
already compared in~\cite{Flynn:2011gf} but we have added more
information available in the literature.}
\begin{center}
\begin{tabular}{|cc|cccc|} \hline
Method               &Ref.& $M_{ccc}$ & $M_{ccb}$ & $M_{bbc}$ & $M_{bbb}$ \\  \hline
Variational Coulomb  & \cite{Jia} & 4760(60)  & 7980(70) & 11190(80) & 14370(80) \\
Variational Cornell  & \cite{Flynn:2011gf} & 4799 & 8037 & 11235 & 14398 \\
Faddeev              & \cite{SilvestreBrac:1996bg}& 4799& 8019 & 11217 & 14398 \\
Bag model            & \cite{Hasenf}& 4790 & 8030 & 11200 & 14300 \\
Quark counting rules & \cite{Bjorken:1985ei} & 4925(90) & 8200(90) & 11480(120) & 14760(180) \\
Const. quark model   & \cite{Roberts:2007ni} & 4965 & 8258 & 11548 & 14834 \\
Const. quark model   & \cite{Vijande} & 4632 & & & \\
Relat. quark model   & \cite{Martynenko:2007je}& 4803 & 8023 & 11285 & 14569\\
Instanton quark model& \cite{Migura:2006ep} & 4773 & & & \\
Hypercentral model   & \cite{Patel:2008mv}& 4736 & 8096 & 11381 & 14451 \\
Sum rules            & \cite{Zhang:2009re}&4670(150) & 7443(150) & 10460(110) & 13280(100) \\
Lattice              & \cite{Meinel} & 4780 & & & 14371(12) \\
Regge estimate       & \cite{Guo:2008he}& 4819(7) & & & \\
\hline
\end{tabular}
\end{center}
\end{table*}

Most computations in the literature are consistent with a triply charmed baryon of around
4800 MeV, and ours will not be different. The $bbb$ baryon is preferred
by most approaches in the range of 14400 MeV. A salient exception is the
sum-rule computation of~\cite{Zhang:2009re} that seems to be
significantly lower.

Closest in spirit to our approach is the Coulombic calculation of Jia~\cite{Jia}, that 
could be considered a Leading Order pNRQCD computation of the static
potential. Indeed, this author employs Eq.~(\ref{LObaryonpotential}),
with parameters specified in table~\ref{tab:paramsJia}. The error bar
quoted in that work corresponds to the author's estimate of higher
orders in perturbation theory.
\begin{table}
\caption{Parameters employed by Yu Jia in an early variational
computation of the triply heavy baryon spectrum with a strong Coulomb potential. The
coupling constant is given at the charm scale (employed in the $ccc$
computation) and the bottom soft scale (employed for all
others).\label{tab:paramsJia}}
\begin{center}
\begin{tabular}{|cc|}\hline
$\alpha_s(0.9\ GeV)$ & 0.59 \\
$\alpha_s(1.2\ GeV)$ & 0.43 \\
$m_c=\frac{M_{J/\psi}}{2}\left(1+2\frac{2\alpha_s^2}{9}\right)$ & $\simeq 1.668\ GeV$\\
$m_b=\frac{M_{\Upsilon}}{2}\left(1+2\frac{2\alpha_s^2}{9}\right)$ & $\simeq 4.924\ GeV$\\ \hline
\end{tabular}
\end{center}
\end{table} 
We will explore the systematics of this computation, extending it in
several ways. First, we will work to two higher orders in perturbation
theory with the potentials now available. Thus, we will ascertain that
this error was underestimated. Second, we will quantify the error
implicit in the one-wavefunction variational approximation (that Jia
also uses) by showing explicit computations in very similar atomic
systems for which the experimental data is available. And third, we will
incorporate a running coupling constant at all steps, and handle the
attending infrared systematic uncertainties by comparing different
methods. The outcome of our work will thus be a much more detailed
understanding of triply heavy baryons in the context of pNRQCD.

\begin{figure*}
	\centering
\includegraphics*[width=4.cm]{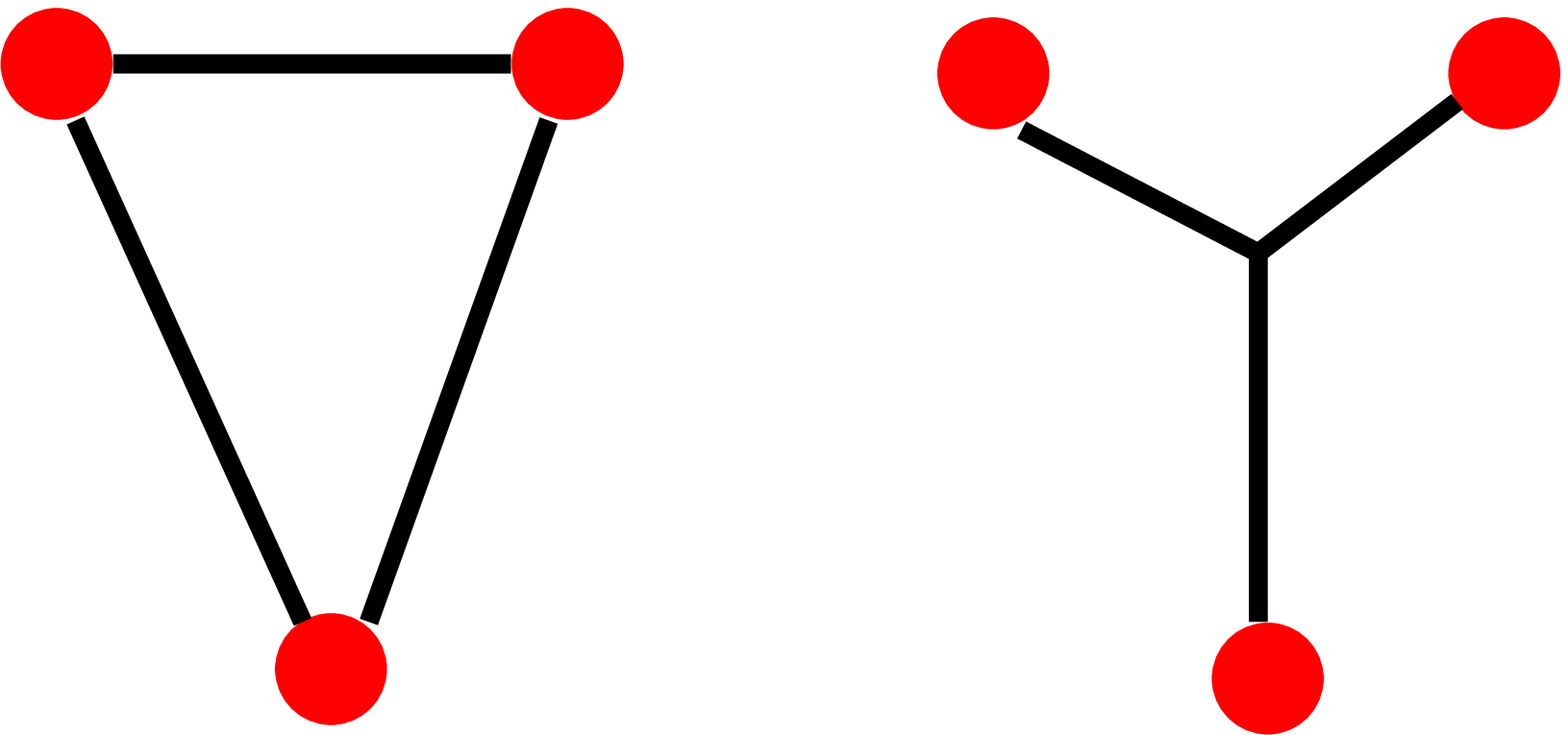}\ \ \ \ \ \ \ \ \ \ \
\includegraphics*[width=8.cm]{FIGS.DIR/ThreebodyFlynn.eps}
\caption{\label{fig:3bodyFlynn}Mass difference between triply heavy
baryons computed with the $\Delta$ type potential (two-body interactions
alone) and 
$\Upsilon$-type potential (three-body interactions alone) reported by
Flynn, Hern\'andez and Nieves in~\cite{Flynn:2011gf}.}
\end{figure*}

It is also worth remarking that in~\cite{Flynn:2011gf} the difference
between the triply heavy baryon mass with a $\Delta$-like two-body
potential and an $\Upsilon$ type potential has been reported in a
variational model computation. We have plotted their results in
figure~\ref{fig:3bodyFlynn}.
As is easily seen, those authors find that the $\Upsilon$ configuration
is slightly heavier, but in any case the difference is only of order
20-40 $MeV$.

This observation is of interest for light-quark baryons, since they can
be more conveniently treated by covariant means, but the (much
simplified) Faddeev equations require a vanishing pure three-body
kernel. That there is not much difference between $\Delta$ and
$\Upsilon$ shape configurations is important information for
establishing that the Faddeev equations are approximately valid, at
least in the heavy quark limit when the soft scale is perturbative~\footnote{Should the soft-scale be non-perturbative, one could obtain some information from other published work\cite{hep-th/9401051,hep-ph/0506065}, but incorporating this in a three-body computation is currently beyond our scope}.

In our perturbative treatment for heavy baryons we will compare the mass
computation with and without the pure three-body force that appears at
NNLO in perturbation theory, finding that this difference is also small,
and thus further reinforcing the conclusion of Flynn {\it et al.}.

\section{Novel computation of triply heavy baryon masses}\label{3bodycomputation}

We treat the 3-body problem in a similar manner
to~\cite{LlanesEstrada:2011jd,Bicudo:2009cr} variationally by employing
a simple wave-function ansatz and computing the expectation value of the
pNRQCD Hamiltonian. The Rayleigh-Ritz variational principle guarantees
that the outcome is an upper bound of the true ground state energy in
the given channel,
\begin{align}
\frac{\la \psi_{\alpha_\rho  \alpha_\lambda} \ar H_{pNRQCD}\ar \psi_{\alpha_\rho  \alpha_\lambda} \ra}{\la \psi_{\alpha_\rho  \alpha_\lambda} \ar \psi_{\alpha_\rho  \alpha_\lambda} \ra} \ge E_0 \ .
\end{align}
The two parameters  $\alpha_\rho$, $\alpha_\lambda$ are then varied to
find the best possible upper bound on energy for the given ansatz. These
two parameters are associated to two momentum-space Jacobi coordinate
vectors, the third independent vector being fixed by the center of mass
condition (hadron at rest)
\begin{align}
k_\rho &= \frac{k_1-k_2}{\sqrt{2}} \\ \nonumber 
k_{\lambda} &= \sqrt{\frac{3}{2}}(k_1+k_2) \\ \nonumber
k_3 &=-k_1-k_2 \ .
\end{align}
We choose as ansatz
\begin{align} \label{ansatzwf}
\psi(k_\rho,k_\lambda)_{\alpha_\rho \alpha_\lambda} = Y_{00}(k_\rho) Y_{00}(k_{\lambda}) 
e^{-k_{\rho}/\alpha_\rho-k_\lambda/\alpha_{\lambda}}\,\,,
\end{align}
which gives reasonable results (we have also checked other forms such as a
rational function). The error incurred in this variational approximation
is estimated below in subsection \ref{sec:errors}. The wave-function in
Eq.~(\ref{ansatzwf}) is symmetrized as needed by invoking it in the
computer programme with different arguments, to guarantee symmetry under
exchange of any two equal quarks. The color singlet wave-function
$\epsilon_{ijk}/\sqrt{3}$ that is implicit in the calculation (and has
already been used in the computation of the color factors of the various
potentials) is then responsible for the antisymmetry expected under
Fermion exchange.

In practice we compute the Hamiltonian's expectation value for the
three-body problem in momentum space.

\subsection{Results in the PS scheme}

With the three parameters $\alpha_s$, $m_c$, $m_b$ in the PS scheme fit
to the meson spectrum, the only sensitivity left to explore is that of
the infrared cutoff scale. In tables~\ref{tab:PS0.4} and \ref{tab:PS0.6}
we present the outcome of the three-body computation in the PS scheme. 

\begin{table} \caption{Ground state triply heavy baryon masses in the PS
	scheme, with infrared cutoff $\lambda=0.4\ GeV$, for various
	orders of perturbation theory. All masses in
	$MeV$.\label{tab:PS0.4}}
\begin{center}
\begin{tabular}{|c|cccc|}\hline
Order & ccc & ccb & bbc & bbb \\ \hline
LO   & 4895 & 8235 & 11535 & 14770 \\
NLO  & 5160 & 8480 & 11750 & 14970 \\
NNLO & 5250 & 8560 & 11805 & 15040 \\
\hline
\end{tabular}
\end{center}
\end{table}

\begin{table} \caption{Ground state triply heavy baryon masses in the PS
	scheme, with infrared cutoff $\lambda=0.6\ GeV$, for various
	orders of perturbation theory. All masses in
	$MeV$.\label{tab:PS0.6}}
\begin{center}
\begin{tabular}{|c|cccc|}\hline
Order & ccc & ccb & bbc & bbb \\ \hline
LO   & 5240 & 8500 & 11640 & 14750 \\
NLO  & 5810 & 9170 & 12460 & 15670 \\
NNLO & 5150 & 8690 & 12100 & 15500 \\
\hline
\end{tabular}
\end{center}
\end{table}

The mass values obtained are significantly higher than in other approaches.
As will be seen in the next section, this is a feature of the PS scheme,
that misses quite some of the binding, as opposed to the pole scheme.
This feeling is reinforced by the observation that the computation with
the lower $0.4 \ GeV$ cutoff does much better, both in terms of binding
and convergence. Particularly bad is the computation with an infrared
cutoff at $0.6\ GeV$ at NLO, that yields an unbelievably high mass.

The results in this and the next subsection satisfy Nussinov's inequalities~\cite{Nussinov}. The first,
\be
M_{\Omega_{bbc}} \le 2 M_{\Omega_{ccb}}- M_{\Omega_{ccc}}
\ee
is a consequence of heavier systems being more bound than lighter systems (as discussed at the end of subsection~\ref{subsec:1overM}). The second inequality,
satisfied by a sizeable amount, reads
\be
M_{\Omega_{bbc}}\ge \frac{M_{\Upsilon}}{2} + M_{B_c}
\ee
and means that mesons are more tightly bound than baryons.  They are well satisfied when the three--body computation is compared to the corresponding two--body computation under the same scheme and conditions employed for parameter fitting.

\subsection{Results in the pole scheme}

\begin{table} \caption{Ground state triply heavy baryon masses in the
	Pole scheme, with infrared freezing point $\lambda=0.4\ GeV$, for
	various orders of perturbation theory. All masses in
	$MeV$.\label{tab:pole0.4}}
\begin{center}
\begin{tabular}{|c|cccc|}\hline
Order & ccc & ccb & bbc & bbb \\ \hline
LO   & 4708 & 7975& 11180 & 14386 \\  
NLO  & 4900 & 8140& 10890 & 14500 \\
NNLO & 4865 & 8150& 11400 & 14683 \\
\hline
\end{tabular}
\end{center}
\end{table}

Tables~\ref{tab:pole0.4} and~\ref{tab:pole0.6} present our results in
the pole scheme with couplng freezing at 0.6 and 0.4 GeV respectively.

Comparing tables \ref{tab:pole0.4} and \ref{tab:PS0.4} we see that the
PS scheme, with its drastic infrared cutoff to avoid renormalons, is
underestimating the binding energy by a large amount of order 300 $MeV$.
To check whether this is ameliorated for yet heavier quarks we have ran
also with $m_Q=10\ GeV$, a quark twice as heavy as the bottom, and with
$m_Q=15\ GeV$.
The results are shown in table~\ref{tab:poleandPS}.

\begin{table}\caption{The difference between the PS scheme and the pole
	scheme seems to persist at masses twice and thrice as big as the
	bottom quark, with the PS scheme underestimating the binding
	energy. Although for asymptotically large masses we believe that
	this difference should ameliorate, we do not see it presently.
	\label{tab:poleandPS}}
\begin{center}
\begin{tabular}{|c|c|c|} \hline
Scheme & Quark mass ($GeV$) & Baryon mass ($GeV$) \\ \hline
Pole   & 5.08 & 14.68 \\
PS     & 5.06 & 15.04 \\
Pole   & 10 & 29.22 \\
PS     & 10 & 29.67 \\
Pole   & 15 & 43.95 \\
PS     & 15 & 44.46 \\ \hline
\end{tabular}
\end{center}
\end{table}

For the 10 $GeV$ quark we obtain a mass of 29.22 $GeV$ in the pole and
29.67 $GeV$ in the PS scheme. Although the difference is now a smaller
percentage of the total mass, it is still very significant in absolute
terms. Therefore we do not expect that a small refitting of parameters,
such as quark masses or coupling constant, will be able to eliminate it
within the present setup.

\begin{table} \caption{Ground state triply heavy baryon masses in the
	Pole scheme, with infrared freezing point $\lambda=0.6\ GeV$, for
	various orders of perturbation theory. All masses in
	$MeV$.\label{tab:pole0.6}}
\begin{center}
\begin{tabular}{|c|cccc|}\hline
Order & ccc & ccb & bbc & bbb \\ \hline
LO   & 4750 & 7950 & 11100 & 14200 \\  
NLO  & 5050 & 8290 & 11470 & 14630 \\
NNLO & 4970 & 8200 & 11340 & 14570 \\
\hline
\end{tabular}
\end{center}
\end{table}

After all numbers have been considered, we deem that the computation
that has the best balance between convergence of perturbation theory and
capture of the infrared physics is that in table~\ref{tab:pole0.4}. To
obtain the best estimate of the physical baryon mass, the results
computed there have to be extrapolated by increasing the binding energy
by 25\% to compensate for the variational approximation (see
section~\ref{sec:errors}).

\begin{figure}
\begin{center}
\includegraphics*[width=7.5cm]{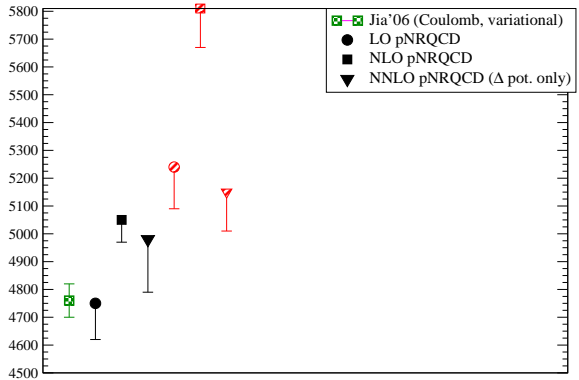}
\end{center}
\caption{$\Omega_{ccc}$ computations in Pole scheme (solid) and PS
scheme (shaded, red online) to LO, NLO and NNLO. The leftmost point
(green online) is the original Coulomb evaluation by Yu Jia with his
quoted error band estimating the NLO effect. Infrared saturation (in the
pole scheme) or cutoff (in the PS scheme) set at 0.6
$GeV$.\label{fig:ccc0.6} The asymmetric error band is our extrapolation
of the missing binding energy due to the variational wave-function.}
\end{figure}

Figure \ref{fig:ccc0.6} compares the results in the PS and pole schemes
with an infrared saturation scale of 0.6 $GeV$, to the three orders of
perturbation theory available.

Several conclusions follow from the figure. First, it is plain that
Jia's calculation is in the right ballpark, but underestimates the
corrections due to higher orders of perturbation theory (note that our
coupling constant is on the low side of the world average, such that a
scheme that will reduce these corrections is hard to imagine). In
addition, one sees that as already mentioned, the PS scheme
underestimates the binding. Finally, and taking into account the
variational error bar (any such calculation underestimates the binding),
the prediction for the $ccc$ mass should be about 4800 $MeV$. We later
will correct this figure up when accounting for the $V_{m^{-1}}$
potential in subsection~\ref{sec:1overm}.

\begin{figure}
\begin{center}
\includegraphics*[width=0.95\columnwidth]{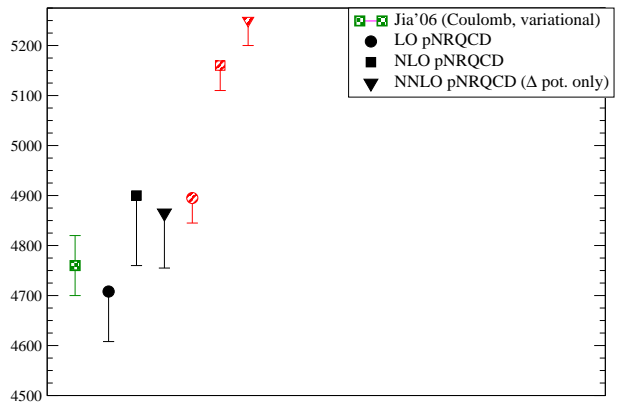}
\end{center}
\caption{Same as in figure~\ref{fig:ccc0.6} but for the IR
saturation/cutoff at a lower scale of 0.4 $GeV$. \label{fig:ccc0.4}}
\end{figure}

Comparing with the results plotted in figure \ref{fig:ccc0.4}, we see
that the region between 0.6 and 0.4 $GeV$ still contributes at least an
additional 100 $MeV$ of binding.

\begin{figure}
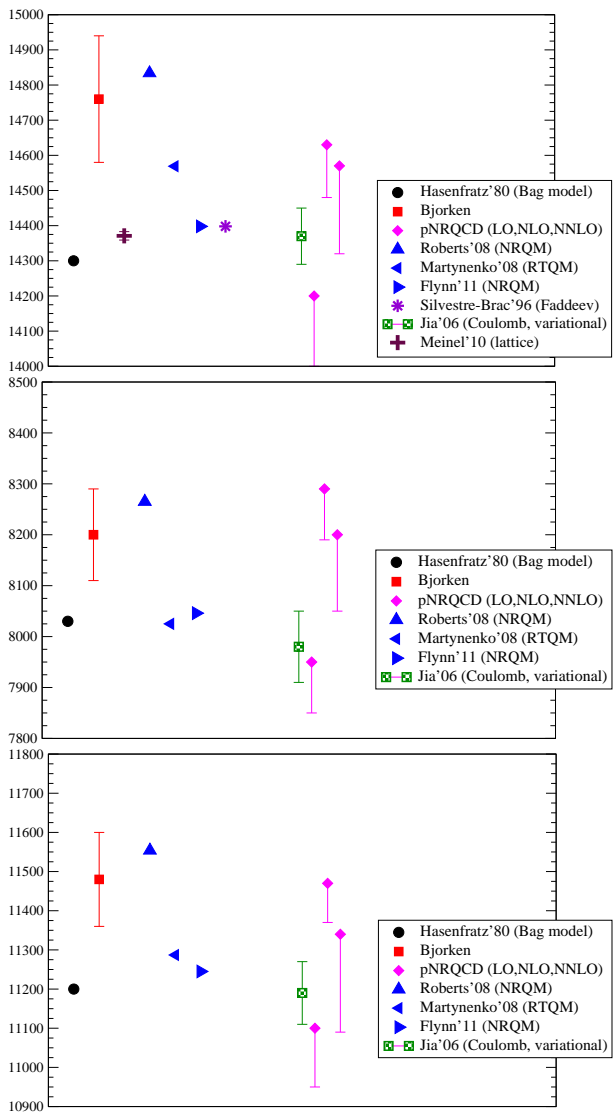

	\centering
\includegraphics*[width=0.95\columnwidth]{FIGS.DIR/AllComputationsbbb.eps}\\
\includegraphics*[width=0.95\columnwidth]{FIGS.DIR/AllComputationsccb.eps}\\
\includegraphics*[width=0.95\columnwidth]{FIGS.DIR/AllComputationsbbc.eps}
\caption{Predictions for the mass of the $\Omega_{bbb}$, and spin
averaged $ccb$, $bbc$ combining results analogous to those of
figure~\ref{fig:ccc0.6} and including other computations as in
figure~\ref{fig:priorcomps}. \label{fig:bbb0.6}}
\end{figure}

Figure \ref{fig:bbb0.6} then plots the predictions for the other
(spin-averaged) triply heavy baryons (in the pole scheme) and gives a
panoramic of other results in the literature.

\begin{figure}
	\centering
\includegraphics*[width=0.95\columnwidth]{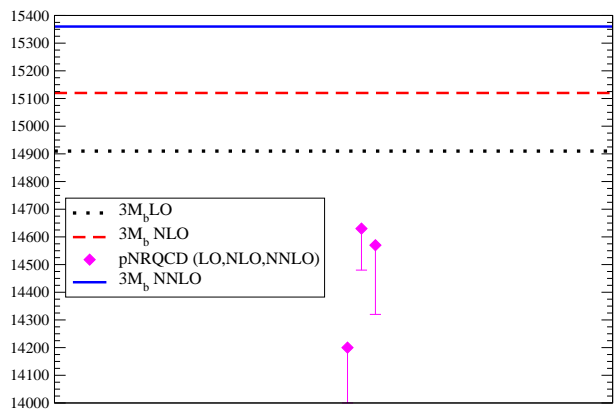}
\caption{Binding energy of the triple-$b$ system. Lines: three times the
pole mass (from bottom to top LO, NLO, NNLO). Symbols: mass of the
$\Omega_{bbb}$.  The graph shows how the resonance mass is better
behaved in perturbation theory than either of the pole mass or the
binding energy. \label{fig:BindingE}}
\end{figure}

Figure \ref{fig:BindingE} shows the size of the binding energy from the
three body variational calculation by comparing it with three times the
pole mass (given that the potential we use is extracted from
perturbation theory, this acts as a dissociation threshold, that should
not be present in a lattice or a Cornell model computation, for
example). It is plain that, although separately the pole mass and the
binding energy do not converge well, there is a cancellation between
them that helps the behavior of the baryon mass in perturbation theory.

\subsection{Effect of the three-body force}

\begin{figure}
\begin{center}
\includegraphics*[width=0.8\columnwidth]{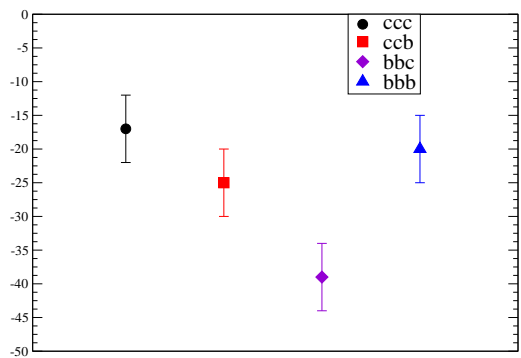}
\end{center}
\caption{\label{fig:three}The effect of adding the three body force to
the NNLO potential is to increase the binding in the amounts visible.
The effect is larger for equal flavor objects (more tightly bound) than
for mixed flavor, and does not depend much on quark masses.}
\end{figure}

Next we address the difference between a computation employing the
intrinsic three-body force, and a computation with only the two body
force.
If we set the scale in Eq.~(\ref{3bodyforce}) according to the ``hard
scale'' prescription
\begin{align*}
\alpha_{s}^3\to \alpha_s(m_c)^3
\end{align*}
we obtain a very small pure three-body contribution of order $1\ MeV$.
If instead, in view of the typical momentum transfer through the three
gluons, we choose the more sensible ``soft scale'' prescription
\begin{align*}
\alpha_s^3\to \alpha_s(q_2) \alpha_s(q_3)\alpha_s(\sqrt{q_2q_3})
\end{align*}
the effect of the (perturbative) three-body force is of order 20-40
$MeV$, in broad agreement with the related (though not equivalent)
estimates of Flynn et al. The result of the computation of the
three-body potential for the different flavor combinations is depicted
in figure~\ref{fig:three}. The effect we find is of size 17 $MeV$ for
ccc, 25 $MeV$ for ccb, 39 $MeV$ for bbc and 20 $MeV$ for bbb, with an
error of about 5 $MeV$ or less.

The immediate conclusion is that in the heavy quark limit, intrinsic
three body forces (defined as those vanishing when one quark is put far
away from the other two) are small in ground state baryons.

\subsection{Effect of the $1/m$ potential}\label{sec:1overm}

\begin{figure}
\begin{center}
\includegraphics*[width=0.5\columnwidth]{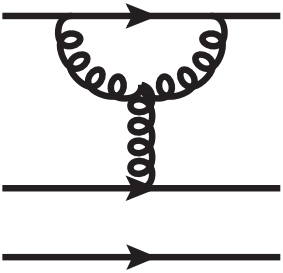}
\end{center}
\caption{Non-Abelian diagram that produces a recoil $V_{m^{-1}}$
potential in triply heavy baryons at NLO. \label{fig:recoilFeyn}}
\end{figure}

Thus far our three-quark results have been based on the purely static
potential. In this section we lift this approximation and study, at NLO,
the effect of adding a $V_{m^{-1}}$ contribution. This recoil correction
has not been worked out in detail in the literature, so we abstain from
attempting an NNLO evaluation. But if we turn to the simplest convention
of~\cite{Brambilla:2000gk}, the NLO $V_{m^{-1}}$ is entirely given by
the non-Abelian diagram with a three gluon vertex, whose equivalent for
baryons is sketched in figure~\ref{fig:recoilFeyn}.

Because the interaction is two-body, the potential in
Eq.~(\ref{1overmpot2}) can immediately be adopted for baryons, with
appropriate kinematics and excepting a color factor. \\
The latter can be worked out easily by noting that the diagram is a
one-loop radiative correction to the quark-gluon vertex on the quark at
the very top of figure~\ref{fig:recoilFeyn}. This non-Abelian vertex
correction is easily seen to increase the usual Gell-Mann color matrix
at the vertex~\cite{Alkofer:2008tt}

\begin{align*}
T^a\to \frac{N_c}{2} T^a \,\, .
\end{align*}
Thus the ratio of color factor in baryons over color factor in mesons is
the same for the static potential as for the recoil correction, to NLO
and in the convention of~\cite{Brambilla:2000gk}
\begin{align}
\frac{C_{3,m^{-1}}}{C_{2,m^{-1}}} = \frac{C_{3,0}}{C_{2,0}} \,\,,
\end{align}
and in practice it is sufficient to divide Eq.~(\ref{1overmpot2}) by a
factor 2 to obtain each of the three possible two-body interactions in
the baryon system.

Next we will show the difference in baryon mass with and without the
$V_{m^{-1}}$ potential. To properly normalize the pole mass and coupling
we first recompute the meson spectrum in section~\ref{fits} and ensure a
best fit shown in table \ref{tab:fitfor1overM}. 

\begin{table*}
\caption{\label{tab:fitfor1overM} Meson observables at NLO  in the pole scheme, with coupling constant freezing only at the very low 0.4 $GeV$ scale. All (spin averaged) masses in MeV.}
\begin{center}
\begin{tabular}{|c|cccc|ccc|}\hline
Potential    & $c\ov{c}(1S)$&$b\ov{b}(1S)$&$b\ov{b}(1P)$& $B_c$ & $\alpha_s(m_Z)$ & $m_c$ & $m_b$ \\ \hline
$V_{m^{-1}}$ & 3068 & 9443 & 9914 & 6200 & 0.107 & 1767 & 5032 \\ 
Static       & 3068 & 9443 & 9857 & 6104 & 0.121 & 1823 & 5093  \\ \hline
\end{tabular}
\end{center}
\end{table*}

To be consistent with the given order in perturbation theory, the
coupling constant runs only at NLO. The quark mass takes an almost
identical renormalization of $-56\ MeV$ (for charm) or $-61\ MeV$ (for
bottom) upon including $V_{1/m}$ that carries over to the baryon
computation and is accounted for in addition to the recoil interaction
there.

\begin{figure}
	\centering
\includegraphics*[width=0.95\columnwidth]{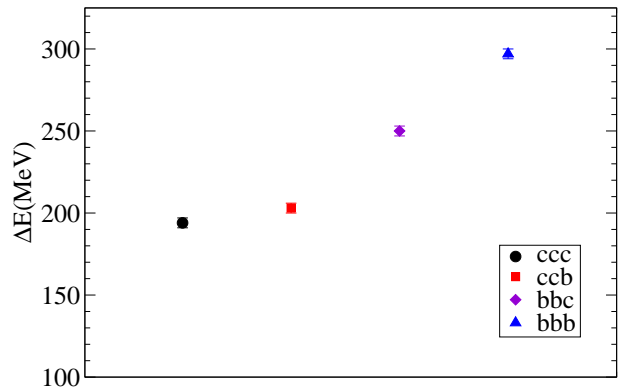}
\caption{Effect of $V_{m^{-1}}$ at NLO from Eq.~(\ref{1overmpot2})
on the ground-state baryon spectrum. Plotted is $M_1-M_0$, the mass
difference including the $1/m$ potential or employing the static
potential alone. The mass of the $\Omega_{ccc}$ can be raised by more
than $150\ MeV$ due to recoil corrections.\label{fig:baryons1overm}}
\end{figure}

The difference between computing baryon masses with the recoil potential
or without it at NLO is depicted in figure~\ref{fig:baryons1overm}.
As can be seen, the effect increases softly from $ccc$ (194(3) $MeV$) to
$bbb$ (297(3) $MeV$) as discussed earlier around
Eq.~(\ref{1overMsurprise}).

\section{Error estimates}\label{sec:errors}
All integrals in the computation of the three-body Hamiltonian matrix
element are evaluated by Monte-Carlo methods and we allow an error of 10
$MeV$ in their computation, except  in our three-body force or recoil
force computations. In those we have demanded an error in the 1-5 $MeV$
range given that we have to subtract two masses. This numerical
uncertainty will be negligible in the final error balance.

\begin{figure}
	\centering
\includegraphics*[width=0.95\columnwidth]{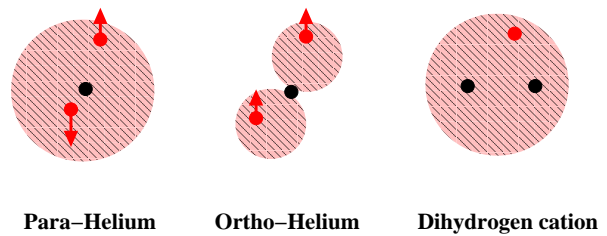}
\caption{\label{fig:dibujomolecular}Three-body systems in atomic physics
that we use to test the variational method and computer programme. From
left to right, para-Helium (one $\alpha$ particle and two electrons
with spin antialigned), ortho-Helium (the two electron spins are now
aligned) and the Hydrogen cation (two protons loosely bound by one
electron alone).}
\end{figure}

To estimate the variational errors we turn to some simple systems in
atomic and molecular physics that can be addressed with the same
techniques, providing in addition a check of the computer programmes. We
take three-body systems made of one electron and two protons (the
dihydrogen cation $H_2^+$), and one $\alpha$-particle binding two
electrons with parallel or antiparallel spins (ortho and para-Helium
respectively). These are depicted in figure~\ref{fig:dibujomolecular}.

Although we are not considering spin interactions, the distinction
between ortho and para-Helium is also important as it checks our
wave-function symmetrization procedure.

We give the matrix elements in terms of reduced momenta $\tilde{k}\equiv
k/(m_e\alpha_{em})$. They read, for atomic Helium
\begin{align} 
\la H\ra_\psi &= m_e \alpha_{em}^2\int \frac{d^3k_1}{(2\pi)^3}\frac{d^3k_2}{(2\pi)^3}\psi^*(k_1,k_2)\times \\
&\bigg[ \frac{1}{2} \left( \tilde{k}_1^2 +  \tilde{k}_2^2 + \tilde{k}_3^2 
\frac{m_e}{m_\alpha}\right) \psi(k_1,k_2) + \nonumber\\ 
&\int \frac{4\pi d^3q}{(2\pi)^3q^2} \Big(
\psi(k_1+q,k_2-q)-2\psi(k_1+q,k_2) \nonumber\\
& \hspace{4.5cm}-2\psi(k_1,k_2+q) \Big)\bigg]\nonumber
\end{align}
and for the diHydrogen cation,
\begin{align} 
\la H\ra_\psi &= m_e \alpha_{em}^2\int \frac{d^3k_1}{(2\pi)^3}\frac{d^3k_2}{(2\pi)^3}\psi^*(k_1,k_2)\times \nonumber\\ 
&\left[ \frac{1}{2} \left( \tilde{k}_1^2 +  \left(\tilde{k}_2^2 + \tilde{k}_3^2 \right) \frac{m_e}{m_p}\right) \psi(k_1,k_2)\right. + \\ 
&\int \frac{4\pi d^3q}{(2\pi)^3q^2} \Big(
-\psi(k_1+q,k_2-q)-\psi(k_1+q,k_2) \nonumber \\
&  \hspace{4.5cm}+\psi(k_1,k_2+q) \Big)\bigg] \,\,,\nonumber
\end{align}
where the $4\pi/q^2$ Coulomb potential is clearly recognizable, and
makes the expression very much alike to our heavy-baryon computation at
LO.

\begin{figure}
\centering
\includegraphics*[width=0.95\columnwidth]{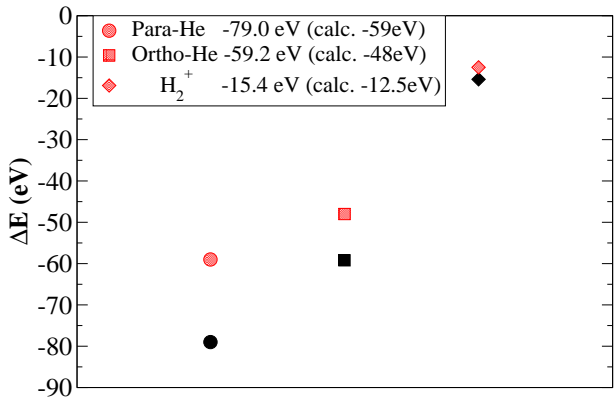}
\caption{\label{fig:moleculecomp}Variational estimate of the various
binding energies in atomic three-body problems, together with the
experimental values.}
\end{figure}

The outcome of these atomic computations is then plotted in
figure~\ref{fig:moleculecomp}. 

From this exercise we estimate the precision of the simple
one-wavefunction, two-parameter variational estimate, to be about 25\%
in error in the computation of the binding energy.  This error can
eventually be reduced to zero by a systematic shell by shell
diagonalization with a large basis, and we are planning such
undertaking, but it exceeds the purposes of the present work.

For now we just note that since the Rayleigh-Ritz variational principle
guarantees that the computed state is less bound than the physical
state, we can reduce the error by the device of increasing the binding
energy that estimated 25\% in the final quoted estimate. This
extrapolation is assisted by our atomic physics computation, since once
the respective scales are pulled out of the matrix elements for heavy
baryons or for atomic Helium, they are very similar.

Returning next to heavy baryons, we observe that perturbation theory
seems to be converging reasonably well, and that, while the jump from LO
to NLO is appreciable, the difference between NLO and NNLO is
substantially smaller and of order $100\ MeV$ at most. This reasoning
applies also to higher order recoil corrections. 

We also incur in a small inaccuracy of order $5-10$ $MeV$ 
in the computation of the $ccb$ and $bbc$ mixed-flavor mesons in employing $N_f=3$ in the appropriate coupling constant, instead of varying the screening $N_f$ with the scale of the various interactions, that may require one of $N_f=3$ or $N_f=4$. This is in order to simplify and speed the execution of computer programmes. In neglecting the charm sea in these baryons we are in line with many other modern computations. $N_f=4$ is correctly set for $bbb$ baryons.

The error is therefore dominated by the treatment of the infrared. The
Potential Subtracted Scheme offers perhaps slightly improved convergence
in perturbation theory, but since the potential is completely truncated
at a low scale, it underestimates the binding energy systematically, and
therefore overestimates the mass by a significant amount (several
hundred $MeV$).

This notorious effect should be absent for infinitely heavy quarks,
where $mv\gg \Lambda_{QCD}$, but since for physical charm and bottom
quarks the scale separation is not very clean, we see that imposing an
infrared cutoff as the PS scheme demands affects the computed binding
energy. 

Turning to the pole scheme with saturated running constant in the
infrared, we see that the binding energy is more in line with other
approaches, and the convergence of perturbation theory is still
acceptable. However the difference between the two approaches advises us
to assign an error of order $200\ MeV$ to the result.

That this infrared sensitivity is caused by the finiteness of the quark
mass can be exposed by the following reasoning. In the case of the
Hydrogen atom, the binding caused by momenta smaller than the inverse
Bohr radius $1/a_0$ is about $13\%$ of the total $-13.6\ eV$. In QCD the
running coupling constant continues growing below $a_0^{-1}$ which is of
order 500 $MeV$, causing a correspondingly higher error. For quarks of
much larger mass however, $a_0^{-1}$ is higher and the low momentum tail
of the wave-function has little overlap with the infrared region where
$\Lambda_{\rm{QCD}}$ influences the result.

\section{Prospects for experimental detection}\label{prospects}

James Bjorken~\cite{Bjorken:1985ei} proposed several decay chains
accessible to experiment that would allow the reconstruction of the
$\Omega_{ccc}$. More recently, Chen and Wu~\cite{ChenWu} have estimated
that $10 fb^{-1}$ of LHC integrated luminosity recorded by a detector
would contain $10^4$ to $10^5$ triply charmed baryons. They further
propose looking for the particular decay chain 
\begin{align*}
\Omega_{ccc}\to \Omega_{sss}\pi \pi \pi \ .
\end{align*}
We agree that this channel provides in principle a very clean signature,
since all four final state particles are charged and can be identified by
$dE/dx$ energy deposition on a tracking chamber. However, the tremendous
combinatorial background that the LHC experiments have to contend with,
often with hundreds of pions in a single event, make the search
extremely difficult.

The reconstruction could be performed also at other experiments, such as
perhaps COMPASS~\cite{Austregesilo:2011jz} and certainly the $B$
factories, operating at less energy and thus with less multi-particle
production. These, particularly COMPASS, are however limited in
statistics and produce less triple charm events. Therefore, it would be
an advantage to combine the spectra in all Cabibbo-allowed four charged
particle channels $\Omega_{sss}\pi \pi \pi$, $\Xi K\pi\pi$, 
$\Sigma KK \pi$, and $p KKK$. 

Still, an alternative route to complete reconstruction of the
$\Omega_{ccc}$ would be to at least measure its mass in a recoil
spectrum.
Since charm is produced by the strong interactions in $c\bar{c}$ pairs,
the $\Omega_{ccc}$ needs to recoil against three charm antiquarks, most
often in the form of three $\ov{D}$ mesons, for example in the reaction
\begin{align} \label{detection}
e^-e^+ \to \Omega_{ccc}\ \ov{p}\ \ov{D} \ov{D} \ov{D}
\end{align}
that, depending on the energy, can be accompanied by any number of pions. 

The interesting feature to exploit is that the $\Omega_{ccc}$ is the
lightest state that can carry three charm quarks. Thus, an alternative strategy
 to reconstructing a decay chain, any of which will have
small branching fractions, would be to search for the recoiling triple
(anti)charm system and recoil antiproton.

Since no triple charm spectrum has ever been published to our knowledge,
searching for three $\ov{D}$ mesons is an interesting undertaking in
itself. This can be accomplished, for example, by a lepton trigger that
tags one of the charm mesons (only 50\% inefficient while suppressing
much background), followed by reconstruction of the other two, or by a
pure hadronic trigger in which all three are fully reconstructed.

The further identification of a recoil antiproton in a small subset of
the events immediately provides an upper bound on the $\Omega_{ccc}$
mass, simply by the missing energy technique against the recoiling
system, even if the $\Omega_{ccc}$ itself was not fully reconstructed.

Baranov and Slad~\cite{Baranov:2004er} estimated that the production
cross section for $\Omega_{ccc}\ov{D}\ov{D}\ov{D}$ at the $Z$-pole in
$e^-e^+$ collisions is of order $0.04fb^{-1}$, which is too small to
have been usable at LEP or SLC. However, we should take into account
that the $\frac{1}{s}$ flux factor allows for a larger cross-section, up
to perhaps 3$fb$, in the $10\ GeV$ region where the $B$ factories
operate. Such cross-section is not unreasonable taking into account that
the B-factories have measured double charm and charmonium
cross-sections. For example, for the very exclusive channel $e^+e^-\to
J/\psi+\eta_c$, Belle finds a cross section of about $26(6)$ femtobarn,
while Babar reports some $18(5)$ femtobarn. Exclusive triple
\emph{charmonium} channels should be another two to three orders of
magnitude smaller, but open flavor channels as the one we propose in
Eq.~(\ref{detection}) will be affected less by wave-function suppression,
if the accelerator reaches sufficient energy.

Adding the masses of all the particles, we find a threshold 
\begin{align}
M_{\Omega_{ccc}} + M_p + 3M_{\ov{D}} \simeq 11440 (250)\ MeV
\end{align}
which lies at a slightly higher energy than Belle's data base at the
$\Upsilon(5S)$, taken around 10860 $MeV$. 

It should be feasible for Super-B and Belle-II to take data at slightly
higher energies around the $\Upsilon(6S)$ resonance and try to identify
a triply charmed spectrum.

As for the detection of an effect of the three-body force in the ground state spectrum, we do not share the optimism in Ref.~\cite{Flynn:2011gf}, since the $25\ MeV$ effect has to be found by comparing experimental data with detailed theoretical predictions of the masses, that, as we have shown, have systematic errors larger by an order of magnitude. 
A cleaner way of extracting this information will have to be devised.

\section{Outlook}

We believe that we have presented a stride beyond the computation of Jia
for triply heavy baryons. In this initial evaluation we have employed
the newly available NLO and NNLO potentials, obtaining results broadly
consistent with other approaches, for the static potential, and also
with recoil corrections.

Although at this point our prediction of the $\Omega_{ccc}$ mass at
$4900(250)\ MeV$ is not particularly precise, it is information stemming
from pNRQCD, the appropriate effective theory of QCD for ground state
triply-heavy baryons, a qualitative improvement over the present,
largely model-based situation. Computations by lattice groups are also underway.

In future work we intend to reduce the uncertainties in this work by
employing an additional, more sophisticated Renormalon Subtracted scheme
(RS) and by attempting a multi-wavefunction systematic diagonalization.
These two undertakings should address the biggest sources of uncertainty
in the present work, the contribution to the binding of the infrared
region, and the variational approximation.

We have also been able to estimate the perturbative three-body force to
be small. This is crucial information for the Faddeev formulation of the
three-quark problem, since three-body forces are totally neglected
there. At least we can now state that the error for the ground state, in
the limit in which all three quarks are heavy, is modest and of order 25
$MeV$ in the mass.

\begin{acknowledgements}
Financial support by grants FPA 2008-00592, FIS2008-01323 
plus 227431, Hadron-Physics2 (EU) and PR34-1856-BSCH, UCM-BSCH GR58/08,
910309, \\ 
PR34/07-15875, SB2010-0012. We thankfully acknowledge the computer resources, 
technical expertise and assistance provided by the CeSViMa and the 
Spanish Supercomputing Network. FJLE is a recipient of a Caja Madrid 
fellowship and thanks the hospitality of the TU-Munich theoretical physics
group and especially  Nora Brambilla and Antonio Vairo, who have been 
influential in several aspects of this work.
\end{acknowledgements}

\appendix

\section{Static potential with BLM scale fixing\label{sec:BLM}}
In this section we introduce the Brodsky-Lepage-McKenzie scale fixing at
NLO (that we have used in practical computations) as well as a sketch on how to proceed at NNLO (that we, however, have not further pursued).
\subsection{NLO potential}

The idea of the BLM renormalization scale choice~\cite{Brodsky:1982gc}
is to absorb the non-conformal terms proportional to the
$\beta$-function of QCD into the running coupling. The resulting scale
$\mu_{\rm BLM}$ is therefore fixed at NLO by demanding that the
$\beta_0$ term cancels.

To one-loop order the idea can easily be carried out by shifting the
scale of the strong coupling constant. This is done by substituting the
solution of the renormalization group equation (with $\beta\simeq
\beta_0$ taken at one-loop)
\begin{align} \label{runningcoupling}
\alpha_s(\frac{1}{\ar{\bf r}\ar^2})&\simeq \frac{\alpha_s(\mu_{\rm BLM}^2)}
     {1+\frac{\alpha_s(\mu_{\rm BLM}^2)}{4\pi} \beta_0 
     \log\left(\frac{1}{\mu_{\rm BLM}^2 \ar{\bf r}\ar^2)}\right) }
\\ \nonumber
&\simeq\alpha_s(\mu_{\rm BLM}^2) + \frac{\alpha_s(\mu_{\rm BLM}^2)^2 }{4\pi}\beta_0
\log(\mu_{\rm BLM}^2 \ar{\bf r}\ar^2) + \dots
\end{align}
Demanding therefore that 
\begin{align*} 
\beta_0(\log(\mu_{\rm BLM}^2 \ar{\bf r}\ar^2)+2\gamma_E     )=0
\end{align*}
we find that
\begin{align} \label{BLMscale}
\mu_{\rm BLM} = \frac{e^{-\gamma_E}}{\ar {\bf r}\ar} \simeq 0.56 \frac{1}{\ar {\bf r}\ar} \ .
\end{align}
For charmonium, one would estimate numerically that

\begin{align*} 
\frac{1}{\ar {\bf r}\ar} \simeq \frac{1}{a_c} = \frac{m_c c^2\alpha_s}{\hbar c} \simeq 0.75\ GeV 
\end{align*}
with the BLM scale correspondingly smaller, about $400\ MeV$ as is usual.

In terms of this scale, the potential to NLO takes the simpler form
\begin{align} 
V^{(0)}_{LO}+V^{(0)}_{NLO} &= -\frac{4}{3} \frac{\alpha_s(\mu^2_{BLM})}{r}\left(
1+ a_1\frac{\alpha_s(\mu^2_{BLM})}{4\pi} \right) \,\,.
\end{align}

\subsection{NNLO potential}

The NNLO contribution to the static potential however continues depending on 
$\beta_0$ and $\beta_1$
\begin{align} 
V^{(0)} &\simeq -\frac{4}{3} \frac{\alpha_s(\mu^2_{BLM})}{r}\left(
1+ a_1\frac{\alpha_s(\mu^2_{BLM})}{4\pi} 
\right. \\ 
&\nonumber+ \left.
\frac{\alpha_s^2(\mu_{BLM}^2)}{(4\pi)^2}\left(
a_2+2\beta_1 \gamma_E + \beta_0^2
\left( \frac{\pi^2}{3} -4\gamma_E^2 \right)
\right)\right)
\end{align}
To eliminate these dependencies one needs to fix the BLM scale at two loops. However the expansion of the running coupling at the next order is significantly more difficult than Eq.~(\ref{runningcoupling}).
Then, to two loops, one has
\begin{align} \label{runningcoupling2}
\alpha_s^{NNLO}(Q^2) &= \alpha_s^{NLO}(Q^2)
-b'\alpha_s^{NLO}(Q^2)^2 \log \left(\log \left(\frac{Q^2}{\Lambda^2}\right)
\right)
\end{align}
where
\begin{align*}
b'= \frac{\beta_1}{4\pi \beta_0}= \frac{153-19N_f}{2\pi(33-2N_f)}\ .
\end{align*}

Since the double logarithm will yield a transcendental equation, one
would have to fix the scale numerically. The computer algorithm proceeds
as follows
\begin{enumerate}
\item Fix the BLM scale analytically at NLO as in Eq.~(\ref{BLMscale}) above.
Then $\mu_{BLM}$ and $\alpha_s^{NLO}(\mu^2_{BLM})$ are given.
\item Obtain $\Lambda$ from Eq.~(\ref{LambdaQCD}).
\item Obtain $\mu_{BLM}^{NNLO\ 2}$ from Eq.~(\ref{runningcoupling2})
by substituting it for $Q^2$ there.
The equation relates $\mu_{BLM}^{NNLO}$ with $\mu_{BLM}^{NLO}$ and the dependency has to be solved  for at the same time that one tries to make vanish all $\beta_0$ and $\beta_1$-proportional terms in the potential.
This is best performed by an iterative Newton's method.
\end{enumerate}
We have found this unpractical for the time being and have only pursued the BLM method at NLO.

\section{Numerical methods}

\subsection{Fourier transform}

To numerically transform potentials between momentum and coordinate space we employ a standard Fast Fourier transform algorithm that implements the discrete formula
\begin{align} \label{FFT}
A_i = \sum_{j=1}^N \exp\left( \frac{2\pi i}{N}(i-1)(j-1)\right) \hat{A}_j \ .
\end{align}

The wanted continuous transform is
\begin{align}
V(r) = \int \frac{d^3q}{(2\pi)^3} e^{i{\bf q}\cd {\bf r}} \hat{V}(\ar {\bf q}\ar)
\end{align}
that after performing the angular integrals and grouping terms becomes
\begin{align}
V(r) = \frac{-2}{(2\pi)^2 r} Re\left(
i\int_0^\infty qdq \hat{V}(q) e^{iqr}
\right)\ .
\end{align}

We discretize a momentum interval $(\epsilon,\Lambda)$ where
$\Lambda\simeq 50\ GeV$ is well above the quark mass scale (the hard
scale) and $\epsilon$ is of order $(20\ fm)^{-1}$, well below any soft
scale in the problems treated. The momentum variable is then stepped
linearly according to $q_j=\frac{\Lambda-\epsilon}{N}j$. The conjugate
coordinate variable is automatically discretized as $r_n = 2\pi
n/(\Lambda-\epsilon)$.

With these choices the vectors that appear in Eq.~(\ref{FFT}) are
\begin{align}
\hat{A}_j &= \frac{(\Lambda-\epsilon)^2}{N^2}j \exp\left(\frac{2\pi i}{N}(j-1)
 \right)\hat{V}(q_j) \\ \nonumber
V_{\Lambda,\epsilon}(r_l) &= \frac{-1}{2\pi^2 r_l} \Re\left(
i\exp(-\frac{2\pi i}{N}l) A_l \right)\ . 
\end{align}

\subsection{Minimization}

We have written a computer programme that employs the well-known Minuit
minimization package from CERN~\cite{James:1975dr} to fix the values of
$\alpha_s$, $m_c$, and $m_b$ with the best possible description of the
observables that we have selected. The Schroedinger equation for the
reduced particle is solved quasi-exactly (on a computer) with the
perturbative potential to LO, NLO and NNLO, for both charm and bottom
quarks. 
This is performed by discretizing the second derivative of the reduced
radial function with the symmetric formula 
\begin{align*}
u''(r_i) \simeq \frac{u(r_{i+1})+u(r_{i-1})-2u(r_i)}{h^2}\ .
\end{align*}
After reducing the Hamiltonian to a numerical matrix, this is
diagonalized. Since the radial problem is one-dimensional, accuracy in
the diagonalization is not an issue.

In the LO evaluation of the potential the coupling constant is fixed at
a renormalization scale $\mu^2=m_c^2$ or  $\mu^2=m_b^2$ employing the
NLO running. Except for this small modification (needed since the same
coupling constant cannot sensibly be used in both charmonium and
bottomonium systems), the computation is perfectly consistent with
perturbation theory, so that at NNLO, in the pure NNLO potential pieces
the LO coupling constant is employed, whereas in the LO piece the NNLO
coupling constant features, and so forth. As far as we can imagine no
contamination is introduced from higher pieces in perturbation theory.
We have also employed the perturbative formulae for $\Gamma_{J/\Psi\to
\eta_c}$.

We have of course checked the sensitivity of the numerical results to
the number of points used in the grid solving Schroedinger's equation (300
turns out to yield very precise answers for low-lying states in the
respective potentials), the maximum size of the grid (that extends to 4
$fm$ and beyond) and other numeric artifacts.

\subsection{Iterative scale determination}\label{app:iterative}

When the argument of the coupling constant depends on the coupling
constant itself such as $\alpha_s=\bar{\alpha}_s(m\alpha_s)$ an
iterative method is in order.

We employ Newton's iterative numerical method. Denoting $\alpha_s^{(n)}$
as the successive approximations to the numeric value, and
$\tilde{\alpha}_s$ the function in  Eq.~(\ref{runningcoupling2}) or
similar, then the coupling constant has been found when
\begin{align}
F(\alpha_s) \equiv \tilde{\alpha}_s\left( \frac{M_{J/\psi}}{4}\alpha_s  \right)-\alpha_s  =0 \ .
\end{align}

Newton's iteration, as long as $F\ne 0$ with sufficient significance, is given by
\begin{align} \label{Newtons}
\alpha_s^{(n)} = \alpha_s^{(n-1)} - \frac{F(\alpha_s^{(n-1)})}{F'(\alpha_s^{(n-1)})}\ .
\end{align}
One can take as initial guess $\alpha_s^0 = \tilde{\alpha}_s(\mu^2)$ at
any standard renormalization scale, and then iterate
Eq.~(\ref{Newtons}).

As an alternative we also employ Jacobi's fixed point method, in which one 
starts with a guess $\alpha_s^{(0)}$ (presumably) larger than the true value, and then iterate the recursive relation $\alpha_s^{(n)} = \tilde{\alpha}_s\left( \frac{M_{J/\psi}}{4}\alpha_s^{(0)}  \right)$ until convergence.

\subsection{Montecarlo computation of three-body matrix elements}

The variational matrix elements in the baryon computation are
multidimensional integrals. Three particles, after center of mass
separation, require six momentum integrations in the kinetic energy
evaluation. Two-body potentials add one loop to the matrix element, up
to nine dimensions. In the three-body force computation there are two
exchanged momenta, and thus twelve-dimensional integrals. We make no
attempt at separating rigid rotations and evaluate all these matrix
elements numerically employing the Vegas
algorithm~\cite{Lepage:1980dq,Hahn:2004fe}.

In the matrix element 
\begin{align*}
\frac{\la \psi \ar H \ar \psi \ra}{\la \psi \ar  \psi \ra} 
\end{align*}
we compute both numerator and denominator (the wavefunction
normalization) numerically. This makes the limits of integration quite
irrelevant for the computation since the function is normalized to one
in the same region where the Hamiltonian matrix element is computed, so
that no probability density is missed. In practice we never extend
integration beyond the hard scale $m_c$ or $m_b$ (one expects the
variational parameters $\alpha_{\rho}$ and $\alpha_\lambda$ to
concentrate the momentum wavefunction around the soft scale $\alpha_s
m_c$ or $\alpha_s m_b$).

We employ a minimum of seven million evaluations of the Hamiltonian and
reach a precision of about 10 $MeV$ for standard computations,
increasing this as needed. A 3 $GHz$ processor can swipe an $8\times8$
set of variational parameters $\alpha_\rho$, $\alpha_\lambda$ in about
an hour.

Our program performs wavefunction symmetrization (or mixed
symmetrization for the $ccb$ and $bbc$ systems) by invoking the
wavefunction with exchanged spins and momentum arguments as needed.
Although here we have not taken spin corrections into account, since
they are unknown for the three-body problem, our program is performing
(trivial) spin sums to allow for a simple upgrade once the spin kernels
for triply heavy baryons become available.


\end{document}